\documentclass[12pt,preprint]{aastex}

\usepackage{graphicx}   

\shorttitle{Quantifying the Luminosity Evolution in Gamma-ray Bursts}
\shortauthors{Kocevski, Liang}

\begin{document}

\title{Quantifying the Luminosity Evolution in Gamma-ray Bursts}

\author{Dan Kocevski \altaffilmark{1} and Edison Liang\altaffilmark{}}

\altaffiltext{1}{Physics Department, University of California, Berkeley, Berkeley, Ca 94709 }
\altaffiltext{2}{Department of Physics and Astronomy, Rice University, Houston, Tx 77005 }

\email{kocevski@rice.edu}
\email{liang@spacibm.rice.edu}


\begin{abstract}
We estimate the luminosity evolution and formation rate for over 900 GRBs by using redshift and luminosity data calculated by Band, Norris, $\&$ Bonnell (2004) via the lag-luminosity correlation. By applying maximum likelihood techniques, we are able to infer the true distribution of the parent GRB population's luminosity function and density distributions in a way that accounts for detector selection effects.  We find that after accounting for data truncation, there still exists a significant correlation between the average luminosity and redshift, indicating that distant GRBs are on average more luminous than nearby counterparts.  This is consistent with previous studies showing strong source evolution and also recent observations of under luminous nearby GRBs.  We find no evidence for beaming angle evolution in the current sample of GRBs with known redshift, suggesting that this increase in luminosity can not be due to an evolution of the collimation of gamma-ray emission. The resulting luminosity function is well fit with a single power law of index $L'^{-1.5}$, which is intermediate between the values 
predicted by the power-law and Gaussian structured jet models. We also find that the GRB comoving rate density rises steeply with a broad peak between $1<z<2$ followed by a steady decline above $z> 3$.  This rate density qualitatively matches the current estimates of the cosmic star formation rate, favoring a short lived massive star progenitor model, or a binary model with a short delay between the formation of the compact object and the eventual merger. 

\end{abstract}


 \maketitle


\section{Introduction} \label{sec:Introduction_ch3}

There are currently roughly three dozen gamma-ray burst events
(GRBs) for which we have independently measured redshifts.  Most of
these redshift determinations come from either identification of
absorption lines in the afterglow spectra, attributed to the gas in
the host galaxy, or from observations of emission lines from the
host galaxy. The combination of these techniques has resulted in a
small but growing GRB sample with redshifts ranging from 0.0085 to
4.5 and a distribution peaking around $z \sim 1$.  From this small
sample, it is already abundantly clear that the isotropic equivalent
energy $E_{iso}$ released in the prompt GRB phase is not a standard
candle. The total radiated energy taken at face value (i.e. when not
correcting for a beaming factor $d\Omega$) clearly spans several
orders of magnitude, ranging from $10^{47}$ for the closest event,
GRB 980425 at $z=0.0085$ \citep{Kulkarni98}, to $10^{54}$ for GRB
990123 at $z=1.6004$ (Kulkarni et al. 1999). Recently
\citet{Sazonov04} and \citet{Soderberg04} have reported on gamma-ray
observations of a nearby underluminous GRB occurring at redshift of
$z=0.106$. These new findings have added to the speculation that
there is either a substantial under luminous population of GRBs
which cannot be seen at large distances and/or that nearby events
($z <$ 0.15) are underluminous compared to distant counterparts,
pointing to the evolution of the average energy emitted by a GRB
with time.

A measure to the extend to which luminosity evolution exists in the
GRB population, along with their true luminosity function and
density distribution, may yield important clues regarding the nature
of gamma-ray bursts and how they're progenitors have evolved with
time.  Although the physics of the underlying GRB engine is hidden
from direct observation and is yet uncertain, the total GRB energy
budget is most likely linked to the mass and/or rotational energy of
the GRB progenitor. Understanding of how this energy budget has
changed with time may offer constraints on progenitor properties and
may ultimately point to the physics leading to their explosions.
Since GRB progenitors are most likely linked to compact objects
(supermassive rotating star, black hole or neutron star mergers)
understanding how the GRB luminosity function evolves with time may
give insight to the host environment in the early universe, namely
the star formation rate or initial stellar mass functions at high
redshifts.

Any attempt at quantifying the evolution of intrinsic source
properties of parent populations must account for Malquist type
biases.  Detection thresholds prevent events below a certain flux
from being observed, resulting in the detection of only bright
objects at large distances.  Combined with the fact that bright
events are typically rare, it is very easy in astronomy to
incorrectly conclude that the distant universe is filled with
extremely bright rare objects.  Any attempt at measuring the
correlation between luminosity and redshift without properly
accounting for selection effects will grossly overestimate the
correlation strength between the two variables. Flux limited samples
are a classic problem in astronomy, which manifested prominently in
early quasar studies. Fortunately, straight forward methods have
been devised to account for such effects based on maximum likelihood
techniques. These methods allow for the correct estimation of the
correlation strength between a truncated data set as well as an
estimate on the underlying parent population. The "catch" of such
techniques is that the overall normalization of the resulting parent
distributions cannot be determined, although their functional forms
are constructed in such a way to account for the data truncations.

These techniques also have to limitation of requiring a large sample
sizes and more importantly, an extremely good understanding of the
survey's detection thresholds (i.e. the flux cutoff for magnitude
limited samples). The use of the current sample of GRBs with known
redshift is limited by both of these restrictions.  The current size
of a little over two dozen bursts does not lend itself well to
producing statistically robust results, especially in the high and
low redshift regimes for which only a handful of events have been
detected.  Furthermore, the sample is an accumulation of
observations from several different spacecraft, all of varying
detector thresholds.  It would seem that these limitations could
only be overcome by the accumulation of a larger data set with
consistent detector thresholds which is expected to come from the
Swift spacecraft and the upcoming GLAST mission.

Fortunately, several authors have announced empirical Cepheid like
correlations linking intrinsic burst properties, such as luminosity
\citep{norris00} and the total radiated energy
\citep{amati02,ghirlanda04a} to other GRB observable.  These
correlations may allow for the determination of burst redshifts directly
from the gamma-ray data, which has the advantage of being relatively
insensitive to extinction and observable at far greater distances
than afterglow line measurements. The first of these correlations
was reported by \citet{norris00}. Using 6 BATSE detected bursts with
known redshift, they found an anti-correlation between the {\it
source} frame lag between the 25-50 keV and 100-300 keV emission and
the absolute luminosity of the GRB.  More recently, \citep{ghirlanda04a} reported an empirical correlation between the collimation correction total energy $E_{\gamma}$ radiated by the burst and the rest frame energy at which most of the prompt radiation is emitted $E_{pk}$.  Using these relationships, it is now possible to estimate "pseudo" redshifts for a much larger number
of GRBs detected by the BATSE instrument which perviously lacked any
information as to their distance.  More importantly, the BATSE
detector threshold is relatively well understood for the entire
sample, making the resulting pseudo redshift data excellent for
statistical analysis. 

In this paper, we examine the issue of luminosity and density
evolution by using a sample of over 900 BATSE GRBs for which the
luminosity and redshift where recently estimated by \citet{band04}
through the use of the lag-luminosity correlation.  We limit our analysis to the lag-luminosity correlation primarily due to the lack of jet opening angle $\theta_{j}$ information that is required for the use of the $E_{\gamma}$-$E_{pk}$ relation.  This relationship requires knowledge of $\theta_{j}$ in order to determine the collimation factor, which is only known for bursts with measured jet break times and hence cannot be used with the BATSE sample in consideration for this paper.  We found that the more general, and much broader, correlation between intrinsic $E_{pk}$ and $E_{iso}$ reported by \citet{amati02} did not provide redshift constraints for a majority of the bursts in our sample.  This is consistent with recent observations by \citet{nakar} and \citet{bandpreece} who also found large fractions of their BATSE samples to be inconsistent with the Amati correlation.  Therefore we limit the current analysis to distances estimated through the use of the lag-luminosity relationship.

To our sample, we apply statistical techniques developed by \citet{Lynden-Bell71} and
\citet{efron92} and first applied to GRB analysis by \citet{lloyd02}
to measure the underlying luminosity and density distribution in a
way that properly accounts for the detection thresholds of the BATSE
instrument.  We find a strong (11.63 $\sigma$) correlation between
luminosity and redshift that can be parameterized as $L(z) =
(1+z)^{1.7 \pm 0.3}$.  The resulting cumulative luminosity function
$N(L')$ is well fit by double power law separated by a break energy
of about $10^{52}$ ergs s$^{-1}$, with the differential luminosity
function $dN/dL'$ exhibiting a power law shape of $L^{-1.5}$ below
this luminosity. We show that the GRB comoving rate density
increases roughly as $\rho_{\gamma}(z) \propto (1+z)^{2.5}$ to a
redshift of $z \approx 1$ followed by a flattening and eventual
decline above $z>3$.  This rate density is in qualitatively
agreement with recent photometric estimates of the cosmic star
formation rate (SFR), as would be expected from massive short lived
progenitors.

In $\S 2$, we describe the data set that we use in our study.  In
$\S 3$ we discuss the statistical methods applied to this data to
estimate the GRB luminosity function and comoving rate density as
well as to test for any correlation between luminosity and redshift.
In $\S 4$ we present the resulting demographic distribution
functions of this analysis followed in $\S 5$ by a discussion of the
implications of the shape and evolution of the luminosity function
and comoving rate density on various jet profile.  We show that
there is no evidence for beaming angle evolution in the current
sample of GRBs with known redshift, suggesting that the variation of
the observed luminosity with redshift can not be due to an evolution
of the collimation of gamma-ray emission.  We conclude by examining
how the similarity between the SFR and the GRB comoving rate density
tentatively favors short lived progenitor models.

\section{Data} \label{sec:data_ch3}

\begin{figure}  \label{fig:data}
\plotone{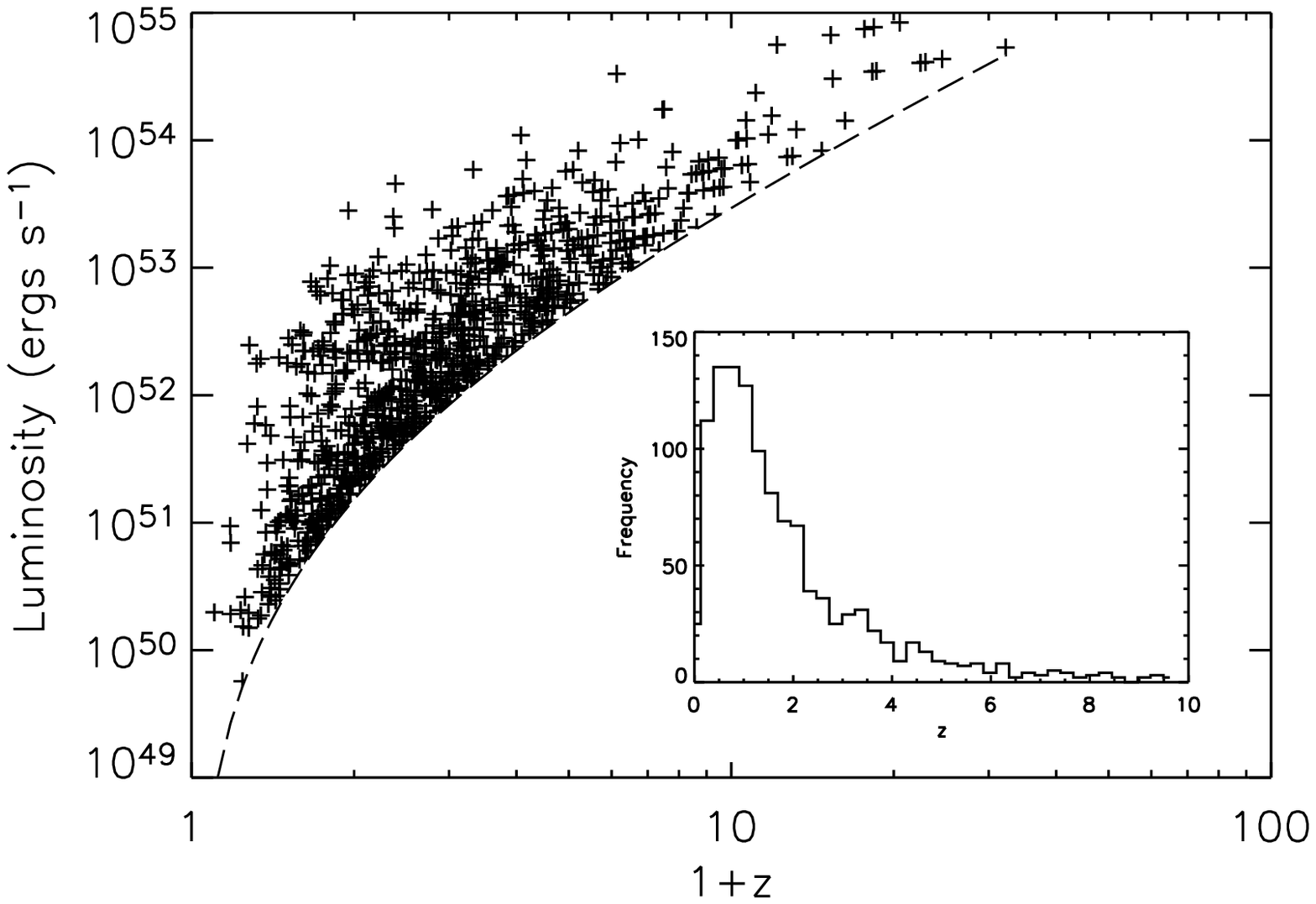}
\end{figure}

For this analysis we utilize data for 1438 BATSE detected GRBs
presented in \citet{band04}, hereafter BNB04. This sample includes
peak photon flux $f_{pk}$ in the 50-300 keV band on a 256 ms
timescale, the burst duration $T_{90}$, and measured lags and their
uncertainties for each burst.  From these lag measurements, the
authors infer each burst's luminosity and redshift by use of the
lag-luminosity correlation, allowing also for an estimation of the
intrinsic $E_{pk}$ and $E_{iso}$ for each burst. Of these 1438
bursts, 1218 have positive lags making them suitable for this
analysis.  This data is shown in Figure \ref{fig:data} with an
imposed flux cut set at 0.5 photon cm$^{-2}$ s$^{-1}$, leaving a
total of 985 bursts.


The lags measurements used in this sample where made using a
cross-correlation analysis similar to that previously employed by
\citet{band93b} and \citet{norris00}. The cross correlation method
has been widely used in x-ray and gamma-ray astronomy, and is well
suited for timing analysis between two signals. In this application,
the normalized discrete cross correlation function is given by
\begin{equation} \label{eq:CCF}
    CCF(\tau)=\sum_{i}^{N-1}\frac{f_{i}(t)*g_{i}(t-t')}{\sigma_{f}\sigma_{g}}
\end{equation}
where $t'$ is commonly referred to as the lag between $f(t)$ and
$g(t)$ and $\sigma_{f}=\langle f(t)^{2}\rangle^{1/2}$. By maximizing
the CCF function (i.e. by maximizing the area of the product of the
two functions) as a function of $t'$, an estimate of the temporal
offset of the two signals can be made. If $g(t)$ leads $f(t)$ by
$t_{0}$ (i.e. $f(t)=g(t+t_{0})$) than the CCF curve peaks at
$t'=t_{0}$.

In BNB04, the authors utilize 64ms count data gathered by BATSE's
Large Area Detectors (LADs) which provide discriminator rates with
64 ms resolution from 2.048 s before the burst to several minutes
after the trigger \citep{fish94}. The discriminator rates are
gathered in four broad energy channels covering approximately 25-50,
50-100, 100-300, and 300 to about 1800 keV allowing for excellent
count statistics since the photons are collected over a wide energy
band. BNB04 measure the temporal offset or lag between channel 3
(100-300 keV) and channel 1 (25-50 keV) light curves to produce the
CCF31 lags listen in their sample.

The shifting of the GRB spectra out of (or into) the observers frame, otherwise known as the k-correction, was accounted for in the analysis performed in BNB04.  They perform spectral fits for most of the bursts in their sample and for those which cannot yeild a fit, a "Band" spectral model with average parameters is assumed for the spectra.  The effects of time dilation and k-correction are then used to obtain the source frame lag and also applied to the energy flux to obtain a bolometric luminosity.

\begin{figure}  \label{fig:laglumplane}
\plotone{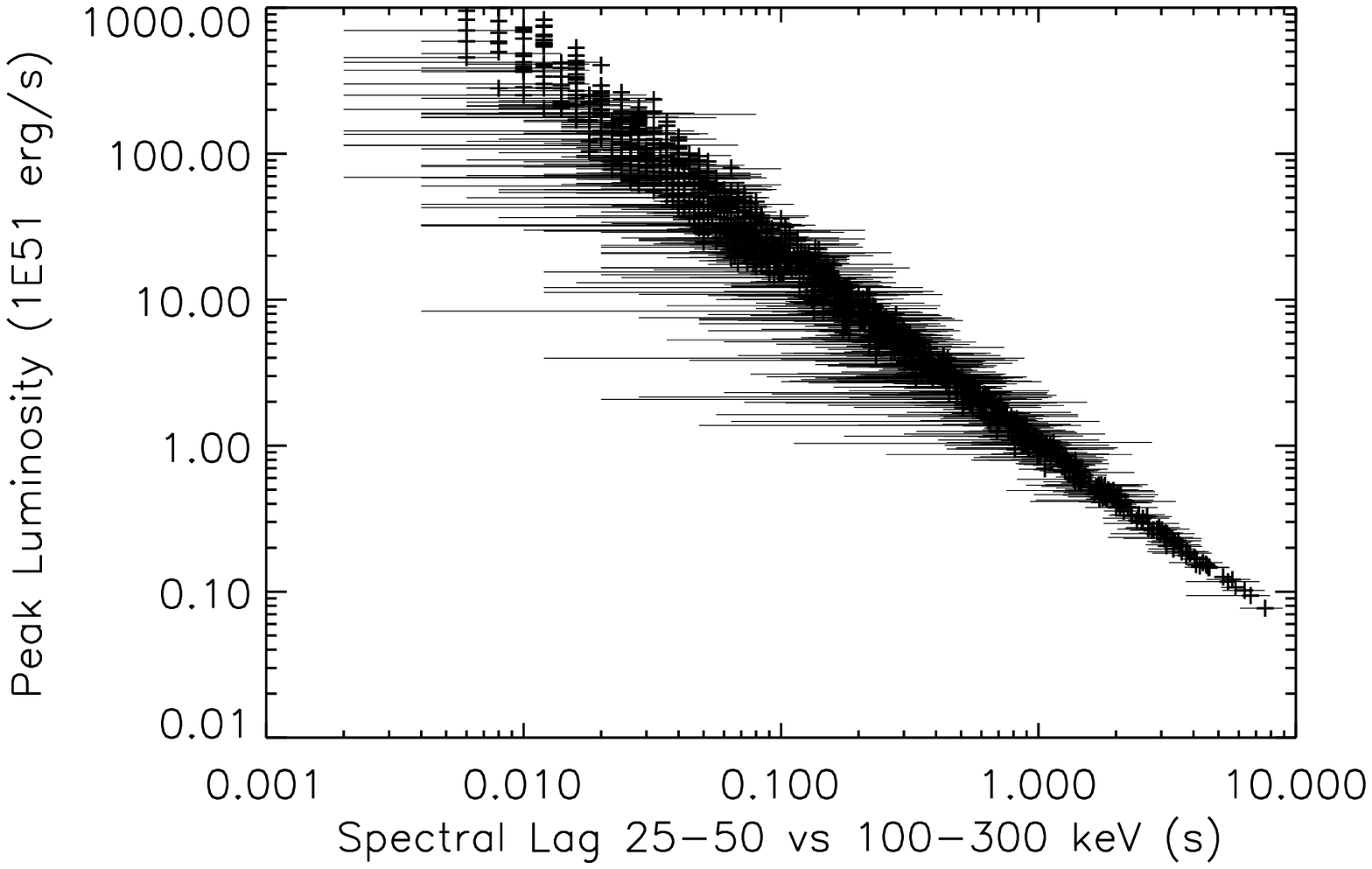}
\end{figure}

As was the case in the original \citet{norris00} paper, the CCF method used in BNB04 can result in lag measurements which are less than the 64ms time resolution of the BATSE instrument.  In these cases, the associated errors of these values tend to be quite large, reducing the significance of their associated luminosity and redshift values.  These errors are taken into consideration in the maximum likelihood techniques performed in our analysis.  Therefore, bursts with extremely short lags (and hence high luminosity's) are weighted accordingly. A plot of the lag-luminosity plane for the events under consideration along with the errors in the lag measurements are shown in Figure \ref{fig:laglumplane}.

\subsection{Estimating Redshifts} \label{sec:redshifts}

Using these lag measurements, BNB04 utilize the lag-luminosity
correlation to estimate the luminosity of each event.  This
empirical correlations was reported by \citet{norris00} who used the
CCF method to measure the lag between BATSE's channel 3 and channel
1 energy light curves for 6 GRBs with independently measured
redshift. They concluded that there was an anti-correlation between
the {\it source} delay in the low and high energy emission and the
absolute luminosity of the GRB showing that high luminosity events
exhibited very small intrinsic (source frame) lag, whereas fainter
bursts exhibited the largest time delay. This empirical correlation
can be expressed as
\begin{equation} \label{eq:laglum}
    L = 2.51\times10^{51} (\Delta t'/0.1)^{-1.15}
\end{equation}
where $\Delta t'$ is the source frame lag related to the observed
lag $\Delta t'_{obs}$ by a time dilation factor of $(1+z)^{-1}$. The
fact that the lag-luminosity correlation relates two source frame
quantities (i.e. luminosity and intrinsic lag) would make it seem
that knowledge of the redshift is needed \emph{a priori}.  As it
turns out this is not the case.  A simple numerical iteration
routine can be used to solve for the redshift of a GRB which lacks
any information as to its distance. This is done by first making an
initial guess for $z$ (say $z \sim 1$) to obtain the lag in the
comoving frame $\Delta t' = \Delta t'_{obs}/(1+z)$.  This in turn
gives us an initial value for the luminosity through the use of the
lag-luminosity relation. This luminosity is then used in combination
with the burst's energy flux to obtain a value for the luminosity
distance $D_{L}$ through the standard relation
\begin{equation} \label{eq:DL1}
    D_{L} = \sqrt{\frac{L/d\Omega}{f_{256}}}
\end{equation}
where $f_{256}$ is the peak flux in the 256 ms timescale and
$d\Omega$ is the beaming factor. This distance is then compared to
the $D_{L}$ that can be calculated directly from the guessed
redshift $z$ by assuming standard cosmological parameters ($H_{o} =
65$ km s$^{-1}$, $\Omega_{m} = 0.3$, $\Omega_{\Lambda} = 0.7$) and
using the expression
\begin{equation} \label{eq:DL2}
D_{L} =
(1+z)\frac{c}{H_{0}}\int_{0}^{z}\frac{dz}{\sqrt{\Omega_{m}(1+z)^{3}+\Omega_{\Lambda}}}
\end{equation}
The value for $z$ is then varied until the luminosity distances
obtained from the two separate methods converge to within some
predetermined precision.

We note that it has been suggested by \citet{salmonson01} and
\citet{norris02} that the lag-luminosity relationship should be a
broken power law in order to accommodate GRB 980425.  This burst was
associated with SN 1998bw and when using the distance to the
supernova, the GRB appears under luminous compared to the other
bursts that fall on the lag-luminosity correlation. In their
analysis, BNB04 note that this break has been suggested to fit a
single point, which may or may not be associated with the SNe event
and hence decide to use a single power law of -1.15.

The physical origin of the lag-luminosity correlation is not
immediately clear.  Fundamentally, this observed lag is due to the
evolution of the GRB spectra to lower energies, so a relationship
between the rate of spectral decay and luminosity is expected
\citet{kocevski03a}. This implies that the mechanisms resulting in
the "cooling" of the GRB spectra is intimately related to the total
energy budget of a GRB or its collimation factor. Other purposed
theories attempt to explain the lag-luminosity correlation as being due
to the effect of the viewing angle of the GRB jet \citep{krm02,
Ioka01}, and or kinematic effects \citep{salmonson00}. In any case,
the use of this correlation is similar to methods used to calibrate
Type Ia supernova luminosities based on the empirical correlation
between their peak magnitude and rate of light curve decay (Phillips
1999). The lack of a clear physical interpretation of these
correlations does not immediately preclude their use in determining,
or refining, luminosity estimates.

\section{Analysis} \label{sec:analysis_ch3}

The luminosity and redshift data calculated by BNB04 gives us an
enormous sample from which to investigate the evolution of the GRB
luminosity function.  As with any cosmological source, it is
important and revealing to understand of how the average luminosity
and density has evolved with cosmic time.  Attempting to do so by
simply measuring the correlation coefficient between the flux
truncated luminosity and redshift data in the BNB04 sample without
properly accounting for the detector selection effects would grossly
overestimate the correlation strength.  This is true whenever an
estimate of correlation is made between two variables that suffer
from data truncations, with the resulting correlation coefficient
representing the truncation itself and not the underlying relation.

There have been several methods developed in astronomy to account
for such selection effects, based largely on maximum likelyhood
techniques (see Petrosian 1992 for a review). In our analysis, we
use a nonparametric statistical technique originally proposed by
\citet{Lynden-Bell71} for applications in flux limited quasar
studies. This so called C-Method has been used successfully to
reconstruct underlying parent distributions for quasars and GRBs
samples by \citet{Maloney99}, and \citet{lloyd02} respectively. The
parent luminosity and redshift distributions which the method
estimates allows for the construction of a GRB luminosity function,
a measure of the number of bursts per unit luminosity, and an
estimate on the comoving rate density, a measure of the number of
bursts per unit comoving volume and time.

The C-Method has two important limitations, or stipulations, to its
use.  First, the truncation limit below which no observations can be
made must be well known. This is not a problem in our case, since
the detector threshold of the BATSE instrument is well understood
and BNB04 quantify the truncation limit of their sample. Secondly,
the parent luminosity and redshift distributions can only be
estimated in a bivariate manner if the two variables are
uncorrelated.  This is a limitation of all nonparametric techniques
which rely on the assumption of stochastic independence. Therefore,
it is necessary to first determine the degree of correlation between
the two variables, in our case luminosity and $Z = 1+z$, and then
produce an uncorrelated data set through the transformation $L
\rightarrow L' = L/g(z)$, where $g(z)$ parameterizes the luminosity
evolution.  Using this uncorrelated data set, it is then possible to
apply the C-Method to estimate the underlying parent luminosity and
redshift distributions.  To estimate the degree of correlation we
use a simple test of independence for truncated data put forth by
\citet{efron92} which is based in part on Lynden-Bell's C-Method.
Below we describe the details of both Lynden-Bell's C-Method and the
Efron $\&$ Petrosian independence test and how they are applied in
our analysis.

\subsection{Test of Independence} \label{sec:efron}

If the variables $x$ and $y$ in a data set are independent, then the
rank $R_{i}$ of $x_{i}$ within that set should be distributed
uniformly between 1 and $N$ with an expected mean $E=(1/2)(N+1)$ and
variance $V=(1/12)(N^{2}-1)$.  It is common practice to normalize
the rank $R_{i}$ such that for independent variables $R_{i}$ has a
mean of 0 and a variance of 1 by defining the statistic
$T_{i}=(R_{i}-E)/V$. A specialized version of the Kendell $\tau$
statistic can be constructed to produce a single parameter whose
value directly rejects or accepts the hypothesis of independence.
This quantity is commonly defined as
\begin{equation} \label{eq:tau0}
    \tau = \frac{\Sigma_{i} (R_{i}-E)}{\sqrt{\sum_{i}V}}
\end{equation}
Based on this definition, a $\tau$ of 1 indicates a 1 $\sigma$
correlation whereas a $\tau$ of 0 signifies a completely random data
set.  See \citet{efron92} for a more detailed (and elucidating)
proof of the applicability of normalized rank statistics.

\begin{figure} \label{fig:associatedsets}
\plotone{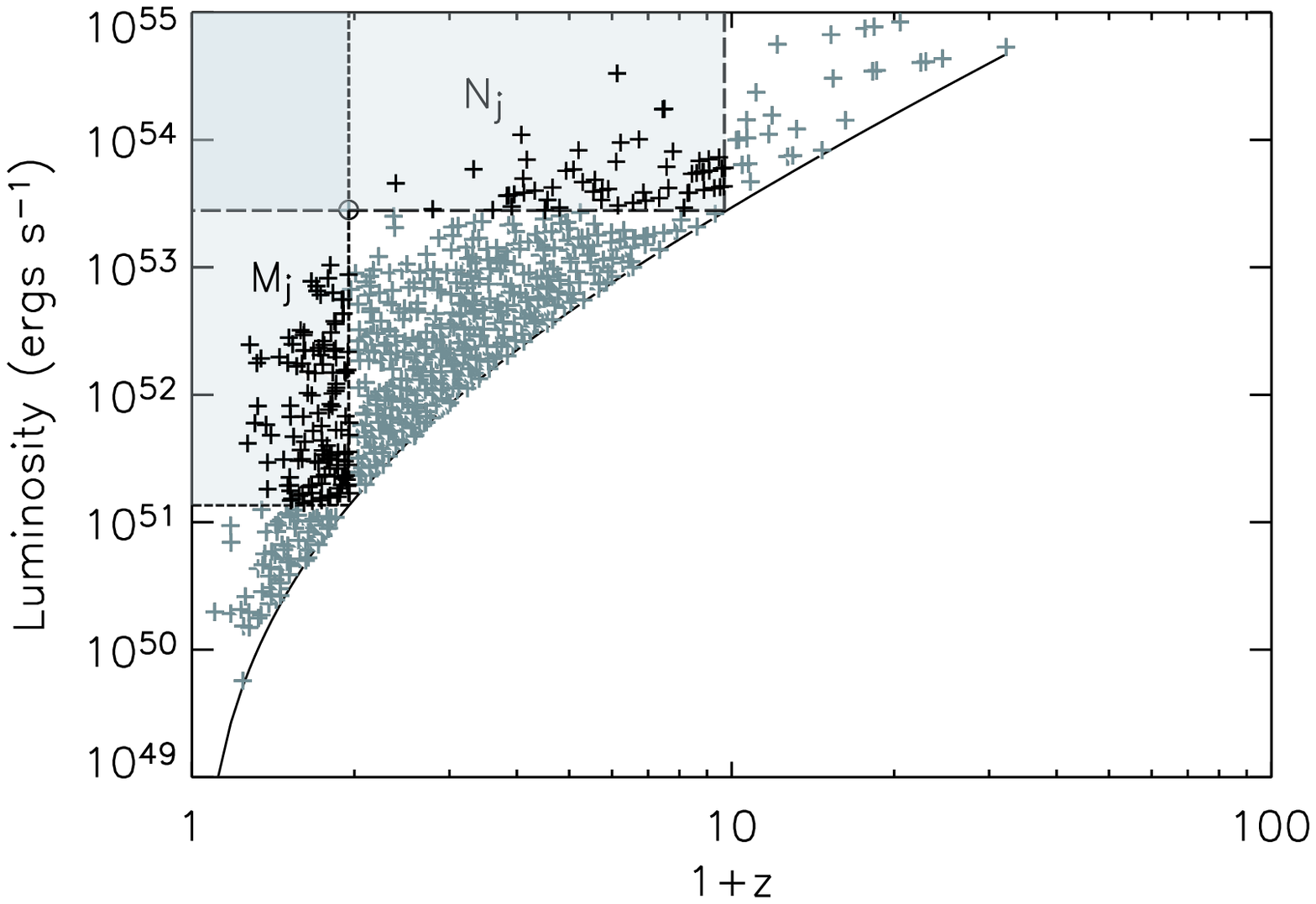}
\end{figure}

The modified version of this rank statistic proposed by
\citet{efron92} to test the independence of truncated data is based
on a simple concept. Instead of measuring the ranks $R_{i}$ for the
entire set of observables, rather deal with data subsets which can
be constructed to be independent of the truncation limit suffered by
the entire sample. This is done by creating "associated sets" which
include all objects that could have been observed given a certain
limiting luminosity. We can define an associated set as
\begin{equation} \label{eq:Ji}
    J_{i} \equiv \{j:L_{j} > L_{i}, L_{lim,j} < L_{i}\}
\end{equation}
In other words, for each burst $i$ there can be constructed a data
subset that includes all events within the range $L_{i} < L <
\infty$ and $0 < z < z_{max}(L_{i})$.  The boundaries of an
associated set for a given burst $i$ are shown as dotted lines in
Figure \ref{fig:associatedsets}.  In this scenario, we expect the rank
$R_{i}$ of $z_{i}$ within the associated set
\begin{equation} \label{eq:Ri}
    R_{i} \equiv \{j\in J_{i} : z_{j} \leq z_{i}\}
\end{equation}
to be uniformly distributed between 1 and $N_{j}$, where $N_{j}$ is
the number of points in the associated set $J_{i}$.  Using these new
ranks, we can again construct the mean and variance, except that now
we replace $N$ with $N_{j}$ such that $E=(1/2)(N_{j}+1)$ and
$V=(1/12)(N_{j}^{2}-1)$. The specialized version of Kendell's $\tau$
statistic is now given by
\begin{equation} \label{eq:tau}
    \tau = \frac{\Sigma_{i} (R_{i}-E_{i})}{\sqrt{\sum_{i}V_{i}}}
\end{equation}
where the mean and variance are calculated separately for each
associated set and summed accordingly to produce a single value for
$\tau$.  This parameter represents the degree of correlation for the
entire sample with proper accounting for the data truncation.  With
this statistic in place, it is a simple matter to find the
parametrization that best describes the luminosity evolution.  This
is accomplished by first choosing a functional form for the
luminosity evolution, which in our case we choose a simple power law
dependence $g(z) = (1+z)^{\alpha}$. We can then make the
transformation $L \rightarrow L' = L/g(z)$ and vary $\alpha$ until
$\tau \rightarrow 0$.

\begin{figure} \label{Fig:fakedata}
\plotone{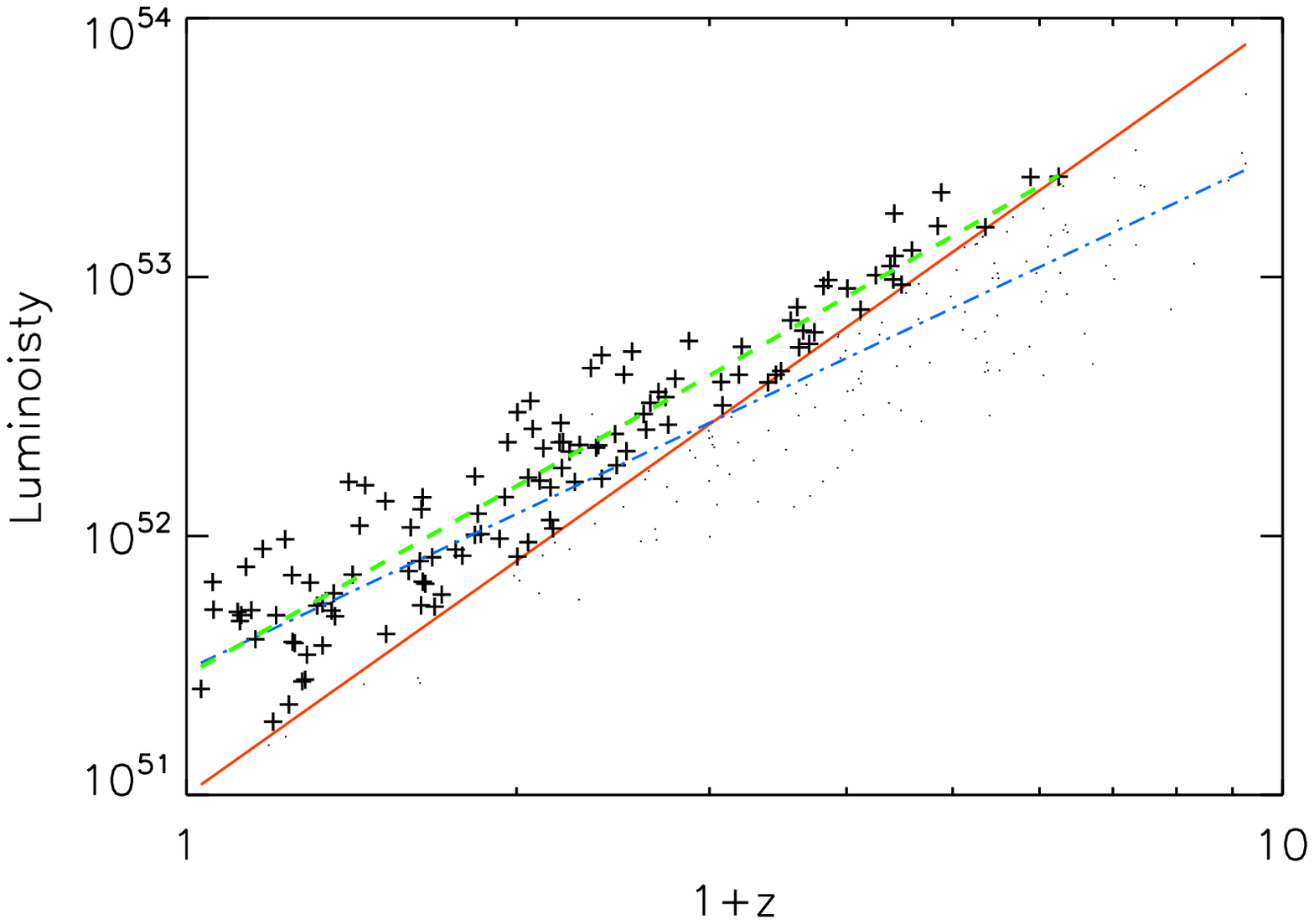}
\end{figure}
\begin{figure} \label{Fig:alphadist}
\plotone{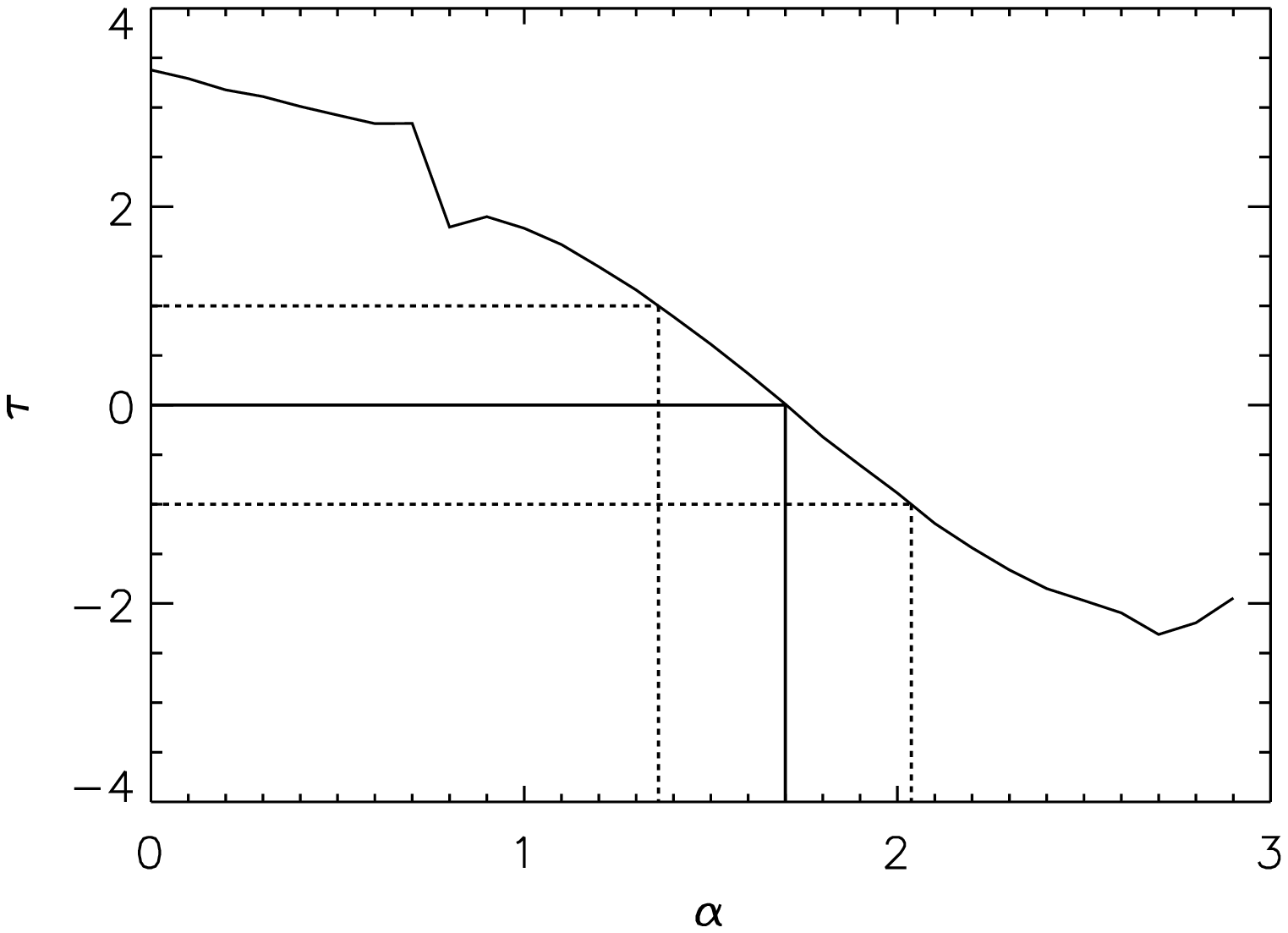}
\end{figure}

An example of how well these methods are able to estimate underlying
correlations in truncated data is shown in Figure
\ref{Fig:fakedata}. Here we have plotted a distribution of fake
luminosity and redshift data with some known power law dependence $L
\propto (1+z)^{p}$ which is subjected to a flux cut $L_{lim} \propto
(1+z)^{q}$ represented by the red dashed line. The crosses show the
observable data whereas the dots represent the data that would
otherwise be undetectable. The long dashed line is the best fit to
the truncated data without any knowledge of the flux cut whereas the
dash dot line is the reconstructed correlation when taking into account
the flux threshold.  This method fails when the undetected data
points become significantly larger than the observable data set,
with the exact boundary at which this occurs depending on the
difference in the power law indices between the underlying
correlation and the flux threshold.  Since these quantities
cannot be known a priori, it is explicitly assumed that a large data
sample contains a sufficient amount of events above the flux
threshold for the method to work.  A histogram of the difference
between the known correlation index and the reconstructed index
$(p-q)$ for multiple such simulations is shown in Figure
\ref{Fig:alphadist}.  The error, or difference between the known $p$
and the measured $q$ is peaked about zero with a fwhm which roughly
matches that error estimates that correspond to the 1 $\sigma$ range
for this parameter given by the condition $|\tau|<1$.

\subsection{Determination of Distribution Functions} \label{sec:distributions}

Once a parametric form that removes the the correlation between $L$
and $z$ is known, it is possible to use nonparametric maximum
likelyhood techniques to estimate the underlying parent luminosity
and redshift distributions.  This luminosity distribution
$\Phi(L_{i})$ represents the cumulative GRB luminosity function with
the redshift distribution $\sigma(z_{i})$ representing the GRB
density evolution. \citet{Petrosian92} has shown that many, if not
most, of the familiar nonparametric methods used in astronomy to
produce $\Phi(L_{i})$ and $\sigma(z_{i})$ reduce fundamentally to
Lynden-Bell's C-Method. Consider the area, or number of events, in
the box produced by the associated set shown in Figure
\ref{fig:associatedsets}. If $N_{1}$ represents the number of points with $L
\geq L_{1}$, then let $dN_{1}$ represent the number of points in the
infinitesimal column between $L_{1}$ and $L_{1}+dL_{1}$. The general
premise behind the C-Method is that if the two variables $(L,z)$ are
stochastically independent, then the ratio between $N_{1}$ and
$dN_{1}$ should equal the ratio between $d\Phi$ and the true
cumulative distribution function $\Phi(L_{1})$
\begin{equation} \label{eq:dnn}
    \frac{dN_{1}}{N_{1}} = \frac{d\Phi}{\Phi_{1}}
\end{equation}
which can then be integrated to find $\Phi(L)$.  In the case of
discrete data points, this integration becomes a summation, yielding
the solution
\begin{equation}  \label{eq:lyndenbell1}
    \Phi(L_{i}) = \prod_{k=2}^{j}\left(1+\frac{1}{N_{j}}\right)
\end{equation}
where $N_{j}$, is the number of bursts in the box defined by
$0<z<z_{max}(L_{j})$ and $L_{j}<L<\infty$.  The value $N_{j}$ is the
same as Lynden-Bell's $C^{-}_{j}$ in that it does not count the
$L_{i}$ object that is used to form the associated set.  Similarly,
we can construct the underlying cumulative redshift distribution
function $\sigma(z_{i})$ by reversing the definition of the
associated set such that $M_{j}$ represents the number of bursts in
the box $0<z<z_{i}$ and $L_{min}(z_{i})<L<\infty$.  Then
\begin{equation} \label{eq:lyndenbell2}
    \sigma(z_{i}) = \prod_{k=2}^{j}\left(1+\frac{1}{M_{j}}\right)
\end{equation}

As mentioned in $\S$ 1, there are several important limitations to
the C-method.  First, the overall normalization of $\Phi(L_{i})$ and
$\sigma(z_{i})$ is arbitrary, so information regarding the absolute
numbers and densities cannot be obtained.  Despite this, the shape
of the bivariate distribution is constructed in such a way that it
accounts for the data truncations. Due to this limitation, all
distributions presented in this paper will have arbitrary
normalizations. Secondly, it is clear from Equation
\ref{eq:lyndenbell1} and \ref{eq:lyndenbell2} that the cumulative
distribution function is not defined when either $N_{j}$ or $M_{j}$
are zero. This limitation restricts the use of the C-method to
samples with a data size sufficiently large to ensure that all
associated sets greater than $j=2$ contain a nonzero number of
points.
\section{Results} \label{sec:results_ch3}

\subsection{Luminosity Evolution} \label{sec:lumevolution}

\begin{figure} \label{fig:taualpha}
\plotone{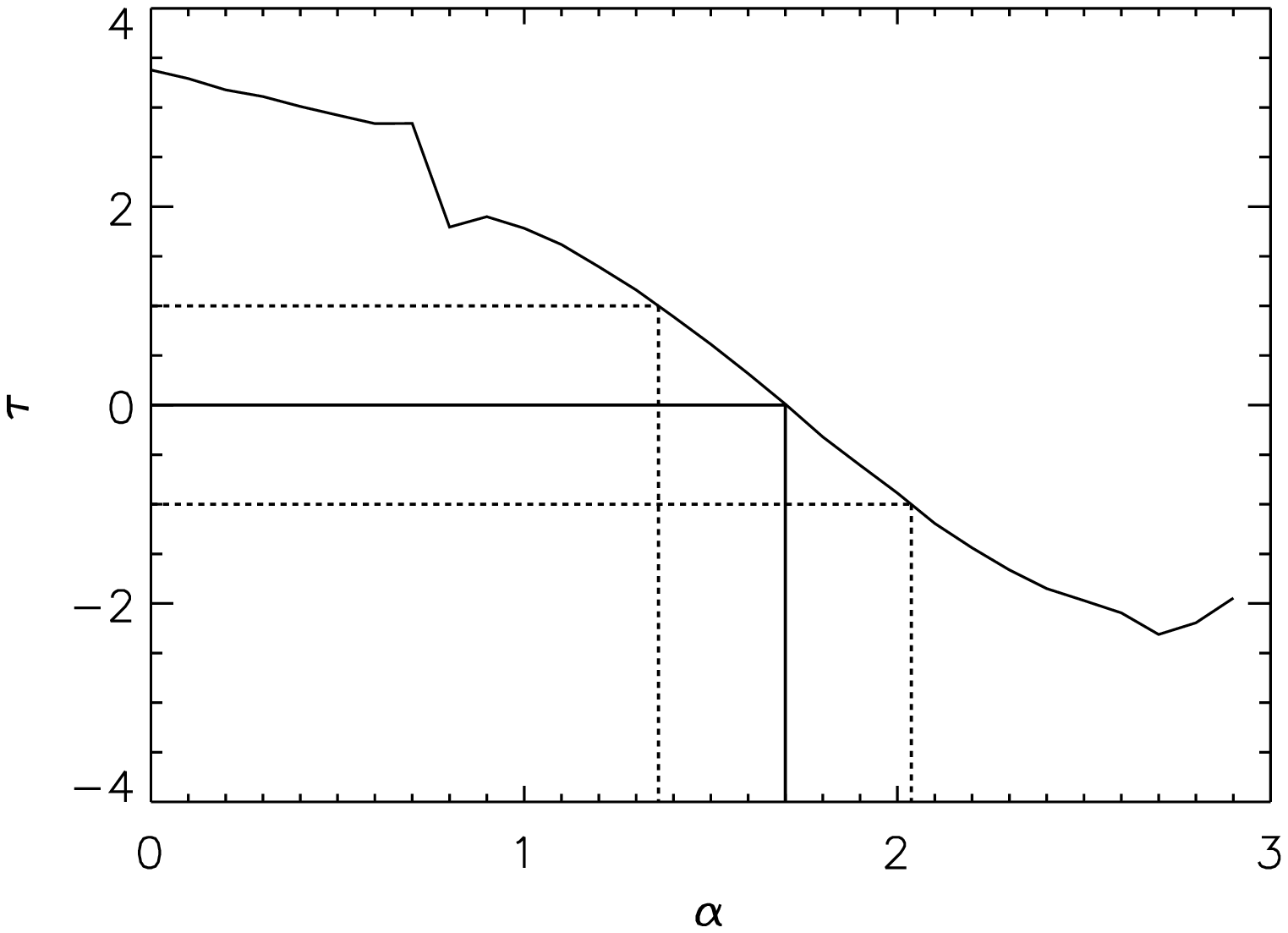}
\end{figure}

We apply the test of independence outlines in the $\S$ 3.1 to the
entire BNB04 GRB sample to test for luminosity evolution. For this
analysis we use the flux threshold suggested by BNB04 of $f_{min}$ =
0.5 photons cm$^{-2}$ s$^{-1}$, decreasing the sample size to 985
bursts.  Applying this method, we find evidence for a strong 11.63
$\sigma$ correlation between luminosity and redshift.  This
evolution is well parameterized by a power law of the form $g(z) =
(1+z)^{\alpha}$, with an optimal value for the power law index (i.e
when $\tau(\alpha)$=0 given the transformation $L \rightarrow L' =
L/g(z)$) of $\alpha$=1.7 $\pm$ 0.3.  The error estimates on $\alpha$
correspond to the 1 $\sigma$ range for this parameter given by the
condition $|\tau|<1$. A plot of $\tau(\alpha)$ vs. $\alpha$ with the
corresponding 1 $\sigma$ levels are shown in Figure \ref{fig:taualpha}.
These findings indicate that the average luminosity (modulo a
beaming factor $d\Omega$) of GRBs in the universe has evolved with
time. Because of the lack of beaming information, it may also be
possible that the luminosity is remaining constant while the beaming
factor $d\Omega$ is actually evolving.  As will be discussed in $\S$
5, there is no observational evidence to suggest that this is the
case.

\begin{figure} \label{fig:alphacut}
\plotone{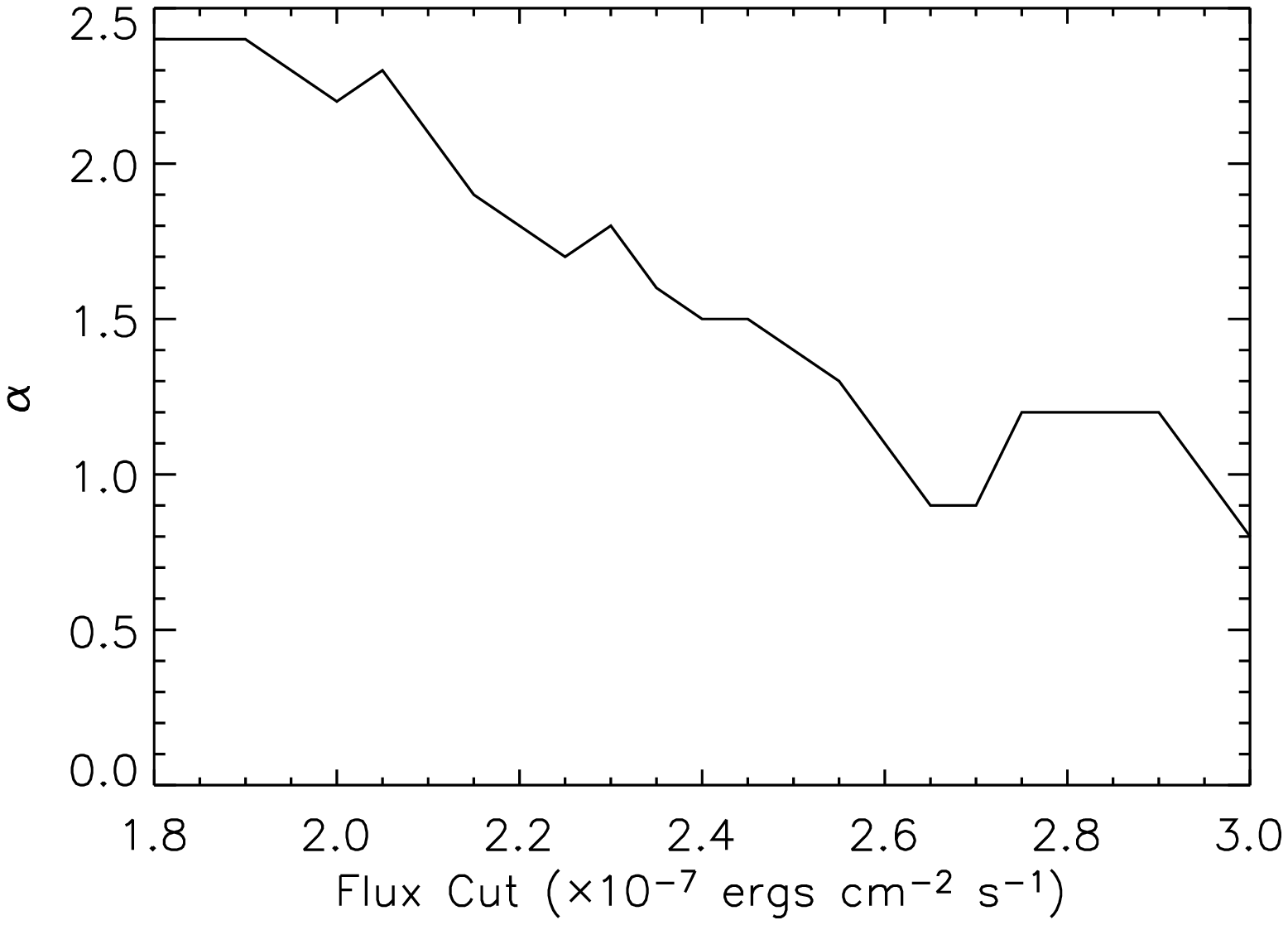}
\end{figure}

It should also be noted that $\tau(\alpha)$ appears to be strongly
affected by the choice of the flux threshold assumed for the sample.
Plotted in Figure \ref{fig:alphacut} is the optimal value for $\alpha$
vs. $f_{min}$. Not surprisingly, if we assume no flux threshold
(i.e. $f_{min}=0$), $\tau$ approaches the overestimated value
received from the standard Kendell $\tau$ statistic.  $\alpha$
similarly approached the value obtained by simply performing a
power-law fit to the truncated data. $\alpha$ decreases steeply with
increasing $f_{min}$, never reaching a stable plateau as one would
hope would happen as the $f_{min}$ approaches the $\emph{true}$
threshold of the detector.  This underscores the importance of
having a good understanding the thresholds of the detector used to
collect the sample.  BNB04 make a strong case for a threshold of
$f_{min}$=0.5 photons cm$^{-2}$ s$^{-1}$ based on where they see a
strong drop off of detected events in the $L-Z$ plane (see their
Figures \ref{fig:alphacut} $\&$ \ref{fig:alphacut}) and we adopt this
value for all analysis presented in this paper.

\subsection{Luminosity Function} \label{sec:lumfunc}

\begin{figure} \label{fig:cumluminosity}
\plotone{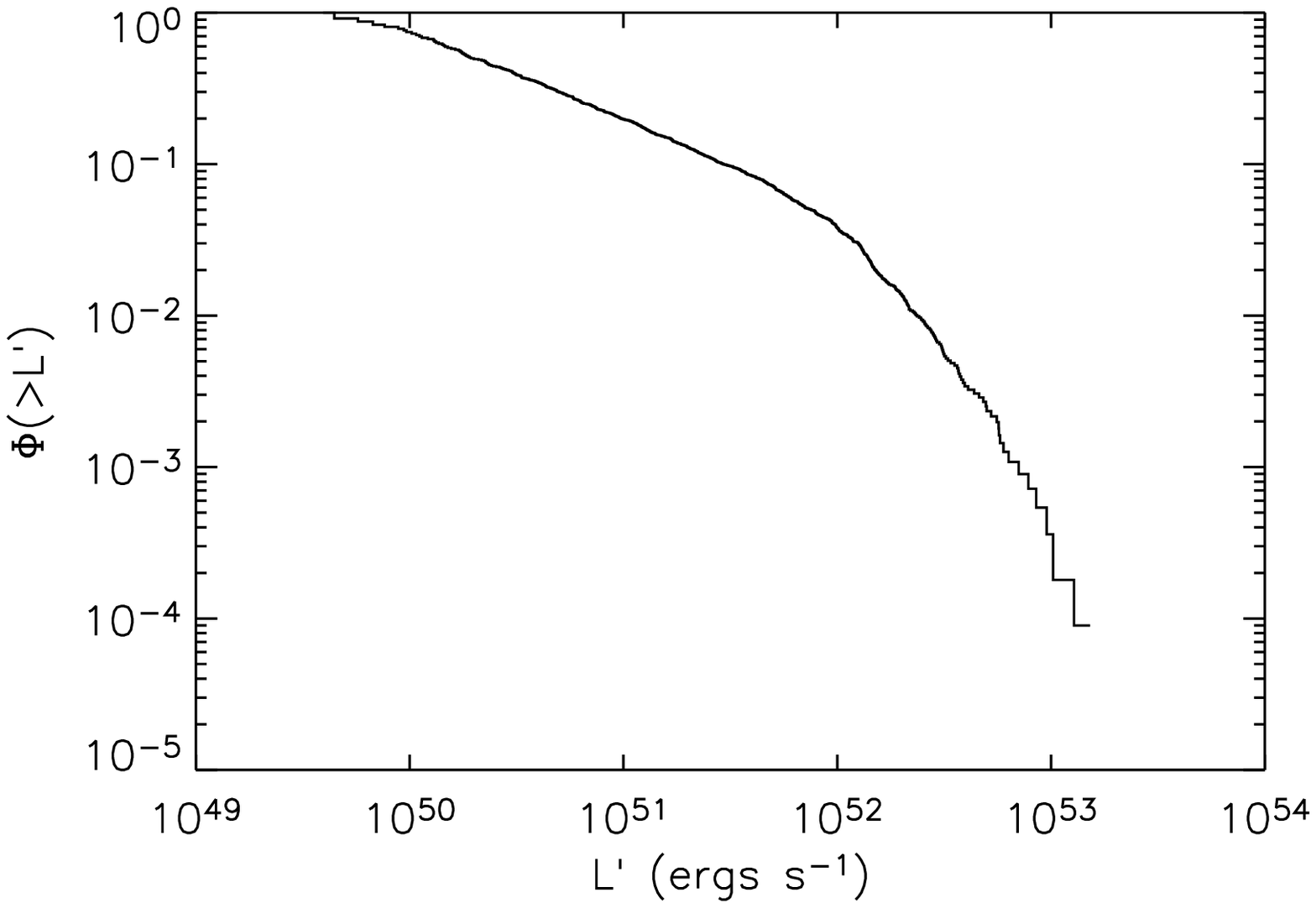}
\end{figure}

The deduced parametric form describing the luminosity evolution
allows us to use the C-method on the uncorrelated parameters $L'$
and $Z$ to obtain the cumulative luminosity function $\Phi(L')$.
Shown in Figure \ref{fig:cumluminosity} is the cumulative $\Phi(L')$
distribution plotted as $\Phi(>L')$ as a function of $L'$ for all
985 bursts. Because the luminosity evolution has been explicitly
removed, this distribution represents the luminosity function in the
present epoch.  Fitting a double power law to the curve yields
$\Phi(>L') \propto L'^{-0.623}$ and $\Phi(>L') \propto L'^{-1.966}$
for the low and high luminosity ranges respectively, separated by a
break at a luminosity of roughly $\sim 10^{52}$.  These slopes are
very similar to those reported by \citet{lloyd02} who found a GRB
cumulative luminosity function with power law slopes of
$k_{1}=-0.51$ and $k_{2}=-2.33$ below and above a break at about
$L'=5.9\times10^{51}$.  These values can also be compared to the
luminosity functions found by \citet{Maloney99} who employ the
C-method to account for selection effects in quasar samples. They
find that the quasar luminosity function exhibits a double power law
form with indices of $k_{1}=-1.16$ and $k_{2}=-3.59$.

\begin{figure} \label{fig:lumfunction}
\plotone{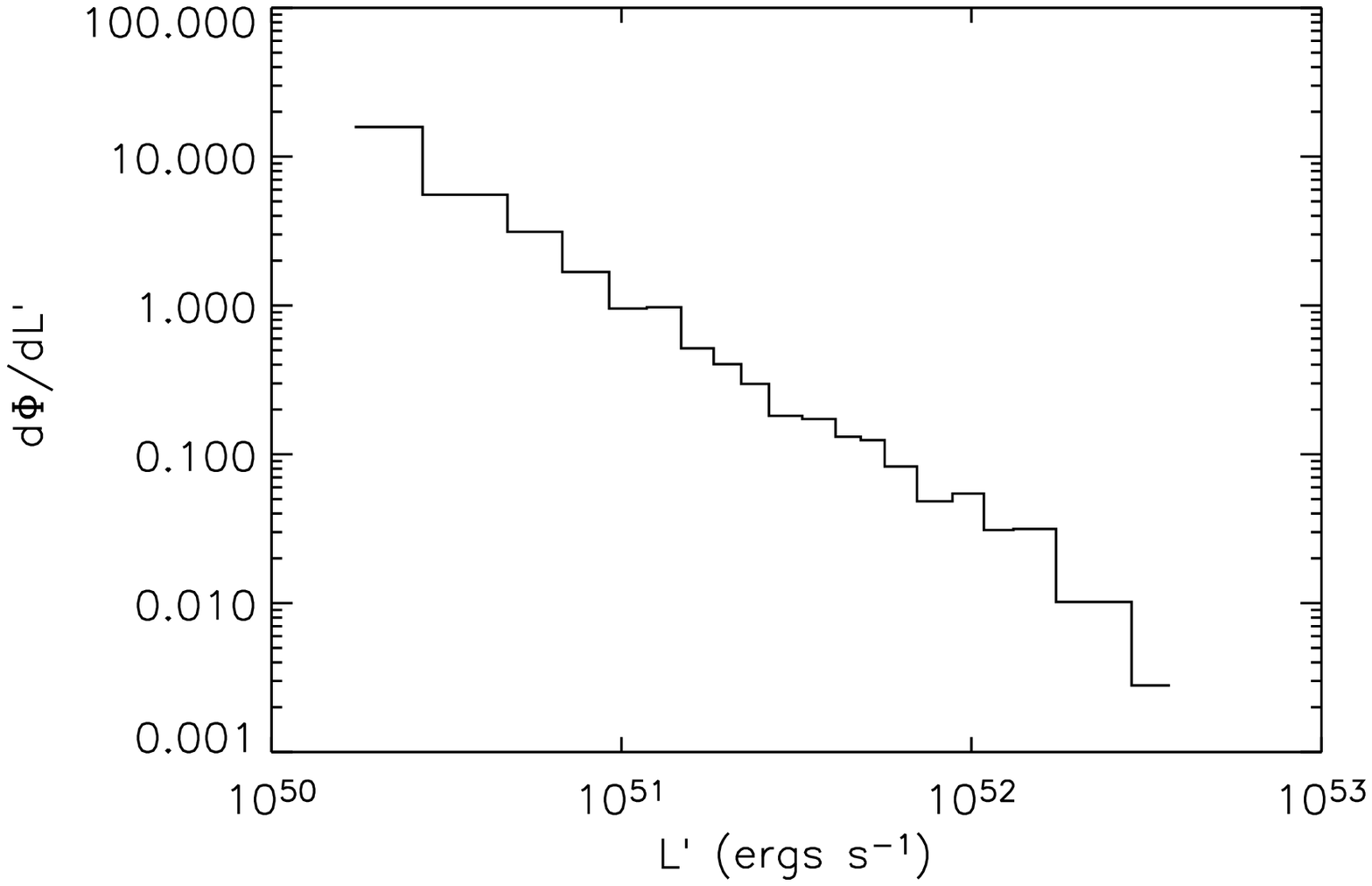}
\end{figure}

Next, we differentiate the cumulative luminosity function with a
3-point Lagrangian interpolation to find the differential luminosity
function $d\Phi/dL'$, or what is commonly referred to as simply the
luminosity function $\psi(L')$.  This function represents the total
number of bursts with luminosity between $L'$ and $L'+dL'$.  A plot
of the $\psi(L')$ vs. $L'$ is shown in Figure \ref{fig:lumfunction}.  The
function falls roughly as $\psi(L') \propto L'^{-1.5}$ below the
break energy of $\sim 10^{52}$ with a sharp decline for higher $L'$.
This power law index is identical to the slope found by
\citet{lloyd02} who found $L'^{-1.5}$ and similar to the index found
by \citet{Schaefer01} who found $L^{-1.7 \pm 0.1}$ from $(L,z)$ data
estimated from a combined use of the lag-luminosity function and
variability-luminosity function, although the latter did not account
for any selection biases in their data set. This value is also
similar to results of several studies that used the measured flux
distribution with an assumed density distribution $\rho(z)$, such as
\citep{Schmidt01} who uses the star formation rate to infer a
$\rho(z)$ and finds $\psi(L') \propto L^{-1.4}$. The shape of the
GRB luminosity function has important implications to jet model
theories which predict specific power law indices for various jet
structures. A comparison between theorized shapes and our deduced
values will be discussed in more detail in $\S$\ref{sec:discussion_ch3}.

\subsection{Density Evolution} \label{sec:densityevolution}

\begin{figure}  \label{fig:cumdendist}
\plotone{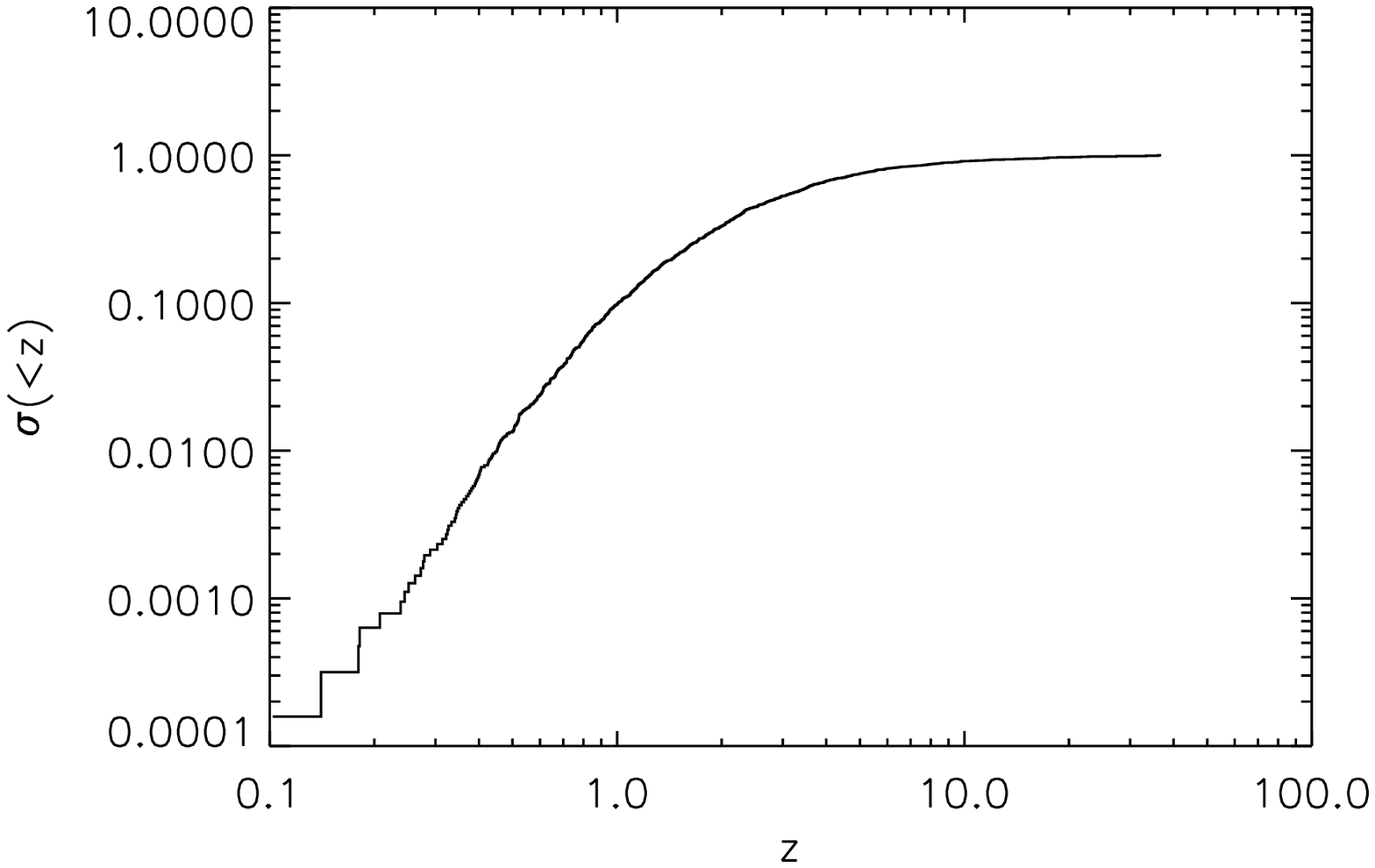}
\end{figure}

\begin{figure} \label{fig:cumdensityvolume}
\plotone{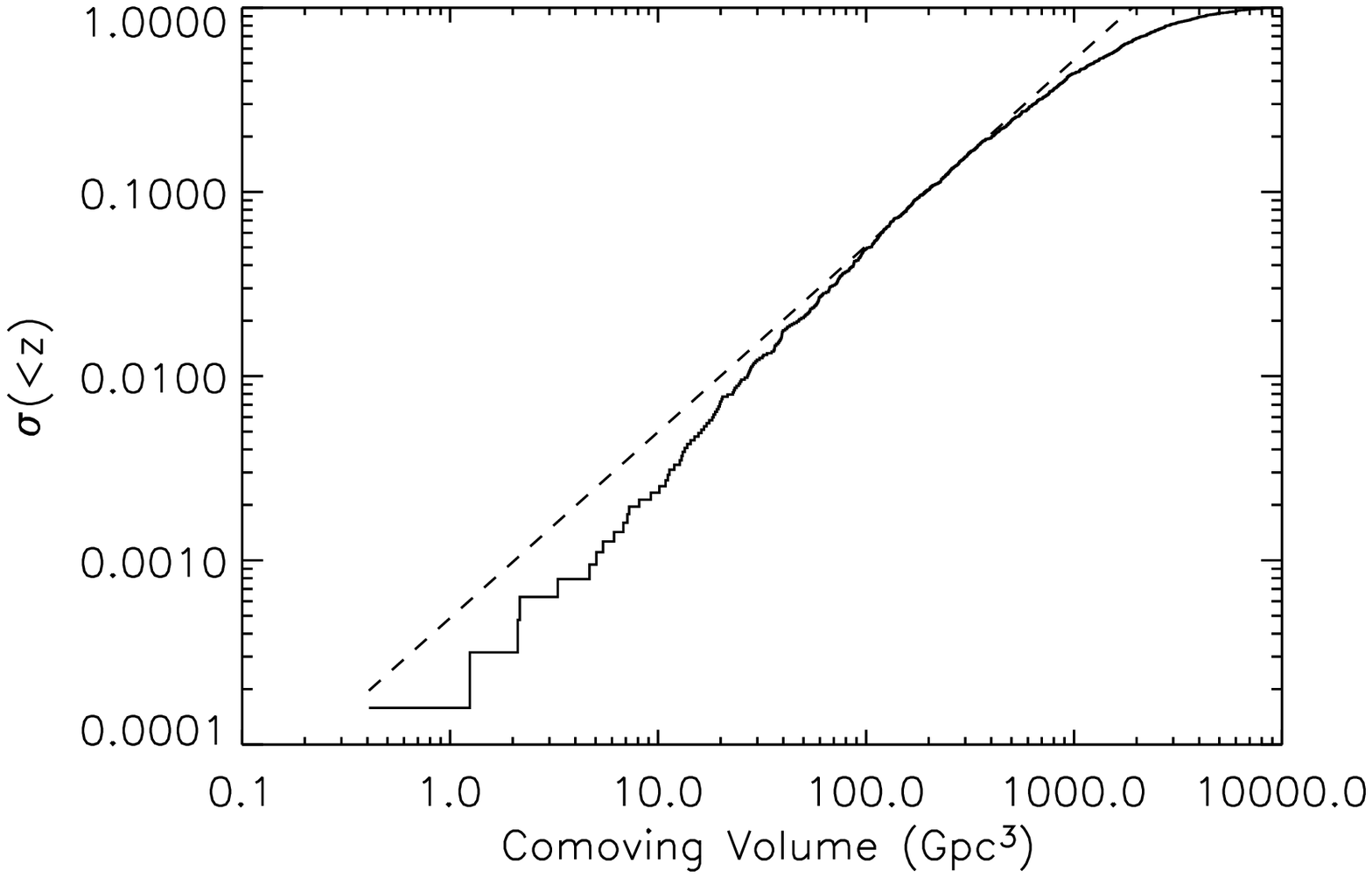}
\end{figure}

Using the alternative definition of the associated set, we can
construct the cumulative density distribution
$\sigma(z)=\int_{0}^{z}\rho(z)(dV/dz)dz$, or the total number of
GRBs per comoving volume, up to a given redshift.  The cumulative
distribution is shown in Figure \ref{fig:cumdensityvolume} plotted as
$\sigma(>z)$ as a function of $z$.  The distribution of GRBs appears
to increase smoothly with $z$, without a pronounced break at any
distance, but with a flattening at high redshift indicating a drop
off of events between $5\leq z \leq 10$.  To get a better look at
this density evolution, we can plot the cumulative density
distribution $\sigma(z)$ as a function of comoving volume $V(z)$ as
seen in Figure \ref{fig:cumdendist} If the density of GRBs per
comoving volume $V(z)$ is constant, i.e. $\rho(z)=\rho_{0}$, then it
should follow that $\sigma(z) \propto V(z)$. We can test for
evolution by fitting $\sigma(z)$ vs $V(z)$ to a simple power law
$\sigma(z) \propto V(z)^{\beta}$ and looking for deviations from the
constant density case.  An index of $\beta \neq 1$ indicates the
presence of density evolution, with $\beta > 1$ and $\beta < 1$
signifying an increasing and decreasing population respectively.
Using the definition of $V(z)$ in a flat universe of
\begin{equation} \label{eq:volume}
V(z) =
\frac{4\pi}{3}\left[\frac{c}{H_{0}}\int_{0}^z\frac{dz}{\sqrt{\Omega_{m}(1+z)^{3}+\Omega_{\Lambda}}}\right]^{3}
\end{equation}
we find that the cumulative density distribution increase with $z$
roughly as $\sigma(z) \propto V^{1.25}$ at low redshifts before
falling off at higher redshifts.  From these results we can deduce
that the GRB density has undergone complicated evolution, increasing
as $\rho \sim V^{0.25}$ before peaking between $z \sim 1-2$ and then
decreasing.  To obtain a more quantitative look at the shape of the
comoving rate density $\rho(z)$, we again use a 3-point Lagrangian
interpolation routine on $\sigma(z)$ to find the differential
cumulative distribution $d\sigma/dZ$.  We can then convert this
differential distribution into a comoving rate density through the
relation:
\begin{equation} \label{eq:rho}
\rho(Z)=\frac{d\sigma}{dZ}(1+z)\left(\frac{dV}{dZ}\right)^{-1}
\end{equation}
In Figure \ref{fig:comovingratedensity} we show the resulting comoving rate
density plotted as a function of $z$.  It can be seen that the GRBs
density function increases out to a redshift between $1\leq z \leq$
then flattens before beginning to show signs of a turn over at a
redshift of $z > 3$.  This is in contrast to previous estimates of
the comoving rate density by \citet{Schaefer01}, \citet{lloyd02},
and \citep{Yonetoku04} all of who find a flattening of the GRB
population with no apparent turn over out to a redshift of $z \sim
10$.  It is also in contrast to results reported by \citet{Murakami} who also used the lag-luminosity correlation to estimate the GRB formation rate.  There the authors find the GRB formation rate increases steadily out to a redshift of at least 4, but it should be noted that this work did not take into account the detector selection effects discussed above so a direct comparison may not be appropriate.  As opposed to these previous findings, the turn over observed in our data quantitatively matches the global behavior of the star formation rate of the universe which has
been observed to peak between $1 \leq z \leq 2$ followed by a steady
decline \citep{madau96,Steidel99}. A more detailed comparison
between the GRB comoving rate density and the supernova and star
formations rates will be continued in $\S 5$.

\section{Discussion} \label{sec:discussion_ch3}

We find an 11.63 $\sigma$ correlation between the luminosity and
redshift data deduced from the lag-luminosity correlation, strongly
suggesting an evolution of the average luminosities of GRBs.  We
show that this correlation can be parameterized as a power law as
$L(z) = (1+z)^{1.7 \pm 0.3}$. This value agrees extremely well with
the results presented in \citet{lloyd02} who found a power law index
of $\alpha = 1.4$ after performing a similar analysis on $(L,z)$
data estimated using the variability-luminosity correlation. These
results imply that the average energy emitted per unit time per unit
solid angle by GRBs was much higher in the distant past compared to
relatively recent events.  This is consistent with previous studies
showing strong source evolution and also recent observations of
under luminous nearby GRBs.  Due to our lack of knowledge regarding
the beaming angle of the bursts in our sample, it is also possible
that the increase in the apparent luminosity is due to an increasing
collimation at higher redshifts. As we will discuss in more detail
below, we find no evidence for beaming angle evolution in the
current sample of GRBs with known redshift and jet opening angle,
suggesting that this increase in luminosity can not be due simply to
an evolution of the collimation of the gamma-ray emission.

\subsection{Comparison to Other Objects} \label{sec:comparisons}

Such a steep luminosity evolution is not uncommon in other
astrophysical objects that show evolution with redshift.
\citet{Maloney99} perform a similar analysis using the statistical
techniques described in this paper on a combination of several
quasar samples and find that the quasar luminosity function evolves
as $L(z) = (1+z)^{2.58}$ up to a redshift of at least 2. There is
evidence that this evolution may then become constant up to a
redshift of at least 3 \citep{Boyle93}.  We find no such break in
the luminosity evolution of GRBs, which in our case can be
adequately fit by a single power law between at least $0 < z < 10$.
The authors also find a density evolution of $\sigma(z) \propto
V^{1.19}$ similar to the power law of $\sigma(z) \propto V^{1.25}$
that we find in GRBs at low redshifts.  A more detailed look at
their comoving rate density estimate shows that the quasar density
rises as $\rho \sim (1+z)^{2.5}$ before peaking at $z \approx 2$ and
then declining rapidly as roughly $\rho(z) ~ (1+z)^{-5}$ for $z
> 2.0$. This is qualitatively similar to the trend we deduce from
the GRB sample, which rises as $\rho \sim (1+z)^{2.4}$ to a $z
\approx 1$ although the proceeding decline is much more shallow as
$\rho \sim (1+z)^{-0.6}$ and extends to at least a redshift of $z
\approx 6$ before dropping off sharply.

There is also evidence for significant evolution in the luminosities
of star forming galaxies, which is perhaps a more relevant
comparison to GRBs because of their suggested association with
active star forming regions \citep{Djorg98}.  Hopkins (2004) used a
compilation of recent star formation rate density measurements as a
function of redshift to constrain the evolving luminosity function
of star-forming galaxies. He finds that the preferred evolution in a
standard cosmology is given by $L(z) = (1+z)^{2.70 \pm 0.60}$ out to
a redshift as high a $z \approx 6$.  At the same time he finds
evidence for a very shallow density evolution given by $\rho(z) \sim
(1+z)^{0.15 \pm 0.60}$, markedly different from steep density
evolution $\rho(z) \sim (1+z)^{2.5}$ that we estimate for GRBs at
low redshift. This would indicate that GRB luminosities have evolved
at a slower rate, but that their density in the past rises much more
steeply compared to the number of star forming galaxies.  It could
also mean that the number of GRBs per star forming galaxy has
evolved rapidly with cosmic time.

\begin{figure}  \label{fig:comovingratedensity}
\plotone{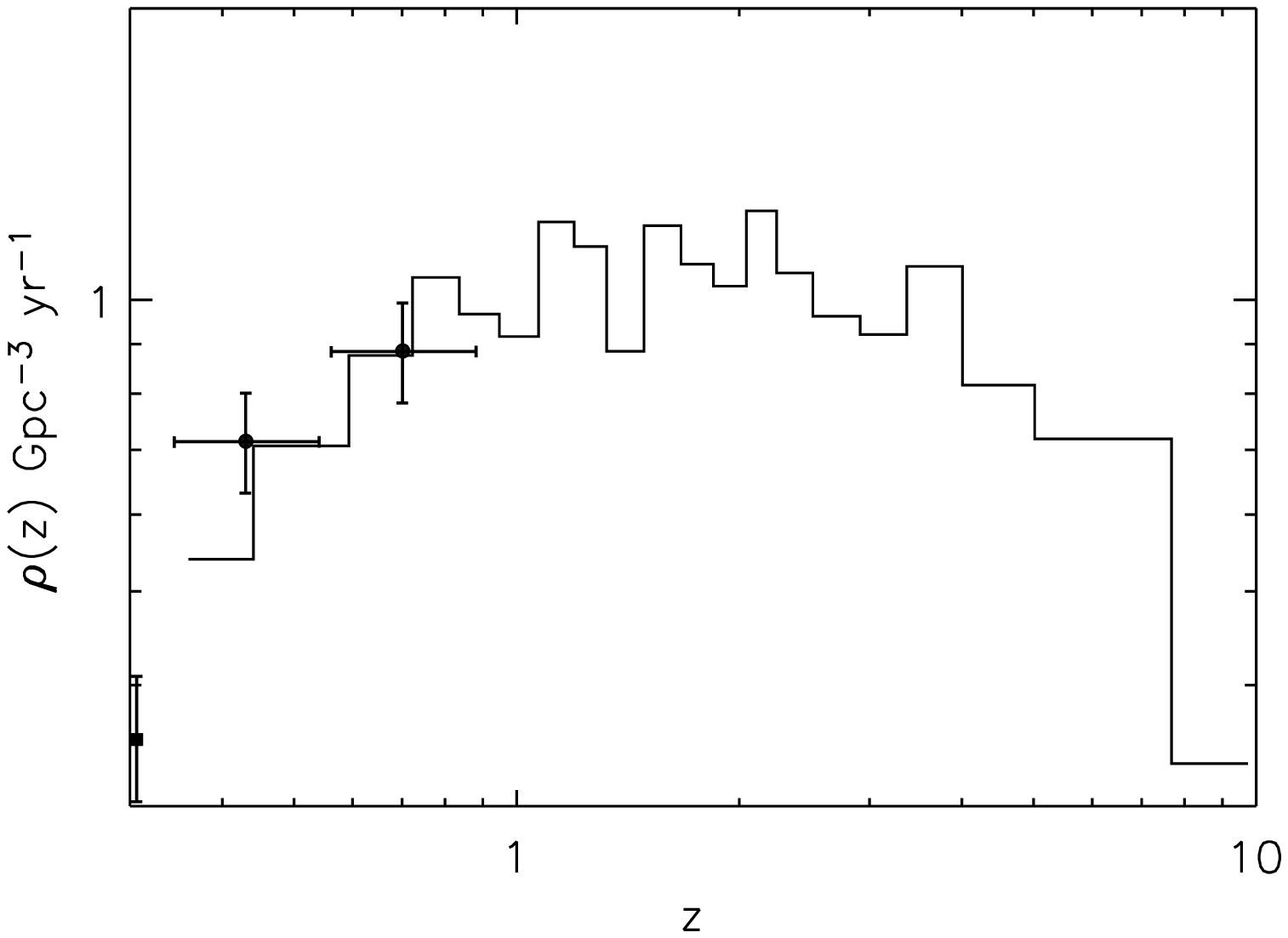}
\end{figure}

Perhaps more interesting is the comparison between the GRB comoving
rate and luminosity densities with the global star formation rate
history of the universe. Because GRBs suffer little extinction and
are potentially detectable out to redshifts of $z \approx 10$, they
could offer a unique tracer to the SFR history. They would allow for
a more complete sampling of dust enshrouded star forming regions
that may be missed in traditional SFR estimates based on the UV
"drop-out" technique that is currently employed to identify Lyman
break galaxies. The shape of the SFR at low redshifts $z<1$ is
relatively well understood, showing an order of magnitude increase
from $0\leq z \leq 1$ \citep{madau96,fall96}. These early estimates
suggest that the star formation activity peaks around $z \sim 1$
followed by a rapid decline at higher redshifts. However, further
observations of hundreds of Lyman break galaxies at redshifts of $z
\sim 3$ and 4 have shown that the SFR may remain constant after
reaching a maximum around $1 \leq z \leq 2$ \citep{Steidel99}.
Recent deep surveys with the Subaru \citep{Iwata03} and Hubble Space
Telescopes \citep{Bouwens03} out to $z \sim 5$ and 6 show evidence
for a mild evolution of the SFR at redshifts $z>3$, with
measurements based on photometric redshifts showing a constant SFR
out to $z \approx 6$ \citep{Fontana03}. These recent SFR estimates
qualitatively match the deduced GRB comoving rate density shown in
Figure \ref{fig:comovingratedensity}. At low redshifts $z < 1$, the GRB rate
density increases as $\rho(z) \sim (1+z)^{2.5}$ roughly matching the
rise in the SFR over the same range, with a peak somewhere between
$1\leq z\leq 2$.  The following flattening and decline between
$2\leq z\leq 6$ in the GRB $\rho(z)$ matches the global properties
of the SFR estimated from the recent deep surveys.


Of course the comparisons between the GRB comoving rate density and
the SFR are simply phenomenological, since we have as of yet no way
of connecting the amount of star formation for a given amount of
GRBs.  Ultimately this conversion factor depends on knowledge of the
GRB progenitor and the initial stellar mass function (IMF) and how
it changes with redshift.  In the case of the collapsar model
\citep{woosley00, macfadyen99}, the rate of GRBs produced for a
given SFR would increase sharply with redshift, as is the case for
all core collapse events, due primarily to the redshift dependence
on the IMF.  However, the connection between the GRB $\rho(z)$ and
the SFR would be straightforward since the progenitors would consist
of massive stars with short lifetimes making them direct indicators
of the SFR at that redshift.  If the mass range of the progenitors
and the redshift dependence of the IMF is know (or assumed) then it
would be possible to calculate a constant that directly relates
$\rho(z)$ to the SFR.  The case is more complicated for binary
merger models since there would be a substantial delay between the
formation of the progenitor star and the final merger event that
produces the GRB.  The distribution in the delay times is not well
known for SNe events, much less GRBs, but it is expected to be large
enough to dissociate the GRB $\rho(z)$ and the active SFR at a given
redshift.  The peak we observed in our deduced values for $\rho(z)$
matching the peak of the current SFR estimates hardly seems
coincidental, tentatively favoring the core collapse models.


It is interesting then to compare our demographic results to that of
various types of supernovae. There is overwhelming observational
evidence and theoretical discussion suggesting a GRBs-SNe
connection, including observations of supernovae bumps in afterglow
lightcurves \citep{stanek03,hjorth} and a deduced collimation
corrected energy that is narrowly clustered around the typical SNe
energy of $10^{51}$ ergs \citep{Frail01,bloom03b}. Although the
intrinsic luminosity of type Ia SNe are \emph{a priori}
\emph{assumed} to be constant with redshift (hence no luminosity
evolution), we can still compare the formation rates of SNe Ia/b/c
to GRBs, although the b/c events are obviously of more relevance to
GRB models. Unfortunately, very little SNe data is available for the
high $z$ universe with only 7 SNe at $z>1.25$ of the 42 SNe detected
in the redshift range of $0.2 \leq z \leq 1.6$ by the Advanced
Camera for Surveys (ACS) on the Hubble Space Telescope
\citep{riess04}. Data on core collapse supernova accounts for only
17 events of this sample, going out to a maximum range of
$z\approx0.7$. \citet{dahlen04} use this data to estimate the core
collapse SNe (CC SNe) rate between $0.3 \leq z \leq 0.7$ and find a
steep (about a factor of $\sim 7$) increase in the SNe
$\rho(z\approx0.7)$ compared to the local rate presented by
\citet{capp99}.  Shown in Figure \ref{fig:comovingratedensity} are their data
points for CC SNe plotted over the GRB comoving density, both rates
normalized to 1 at $z=0.7$.  The two data points, although limited,
do agree with the rise of the GRB $\rho(z)$.  A direct SNe-GRB
comparison at higher redshifts will have to wait until the launch of
the SNAP spacecraft which is predicted to find thousands of
supernovae, including a significant number at high redshift.

\subsection{The Nature of the Luminosity Evolution} \label{sec:lumevolution2}

The observed luminosity evolution that we observed in the $(L,z)$
data leads to the conclusion that the GRB progenitor population has
most likely evolved in such a way as to create more energetic or
more narrowly beamed bursts in the distant past. Speculations on the
nature of this evolution are dependant on the progenitor model and
how the properties of their population are affected by the
conditions of the early universe. In the case of highly rotating
massive stars (i.e. the collapsar model), the overall progenitor
mass and/or rotation rate could be the determining factor. There is
ample evidence suggesting that the so called population III stars
were much more massive than their present day counterparts.  This is
suggested by recent work showing that the stellar initial mass
function (IMF) has evolved with time, having a much higher value in
the distant past. This higher IMF is due to various factors,
although it primarily is due to the lower metallicity in the early
universe.  The amount of material lost to stellar ejecta has also
been shown to be dependant on the stellar metallicity, causing these
early stars to retain more of their mass until their eventual
collapse.

Although the relationship between progenitor mass and emitted energy
and/or beaming angle is not straight forward, there are reasons to
think that this increase in average mass could result in an increase
in the total energy budget available to a burst. \citet{macfadyen99}
show that under the right conditions, the collapsar model could
produce more energetic bursts with increasing stellar mass, up to
some limit dictated by the energy needed by the GRB jet to punch
through the stellar envelope.  Proponents of black hole accretion
disk models have also shown that the rate of accretion onto the
central engine of the GRB increases dramatically as a function of
the progenitor mass, increasing the overall energy available to the
burst.

Unfortunately, a simple increase in the overall energy budget cannot
by itself explain the deduced luminosity evolution.  \citet{Frail01}
and \citet{bloom03} have recently shown observational evidence
suggesting that the collimation corrected GRB energies $E_{\gamma}$
are actually narrowly cluster around the $10^{51}$ ergs typical of
SNe explosions. They come to this conclusion by correcting the
observed prompt isotropic equivalent energy release $E_{iso}$ of
several GRBs with known redshift by a factor of $1 - \cos
\theta_{j}$, where $\theta_{j}$ is the canonical jet opening angle.
These angles are derived from broadband breaks observed in the
afterglow light curves attributed to the slowing of the GRB jet to
the point where the relativistic beaming angle of the radiation
$1/\Gamma$ becomes greater then $\theta_{j}$. \citet{bloom03}, using
a larger sample of bursts, show that the corrected energies cluster
around $1.33\times10^{51}$ ergs with a variance of 0.35 dex, or a
factor of 2.2.  \citet{guetta03} has reported a similar result when
correcting for the isotropic luminosity $L_{iso}$, although not as
narrow as the $E_{\gamma}$ results. If the collimation corrected
energy and luminosity are indeed invariant with redshift, it
directly implies that the brightening of the apparent isotropic
equivalent luminosity is actually due largely to an increase in the
beaming factor as a function of redshift and not an increase in the
overall energy of the burst. There are physical arguments that can
be made in the case of the collapsar model that would suggest that
more massive progenitor stars could indeed produce more collimated
jet outflows.



\begin{figure} \label{fig:fluencevsredshift}
\plotone{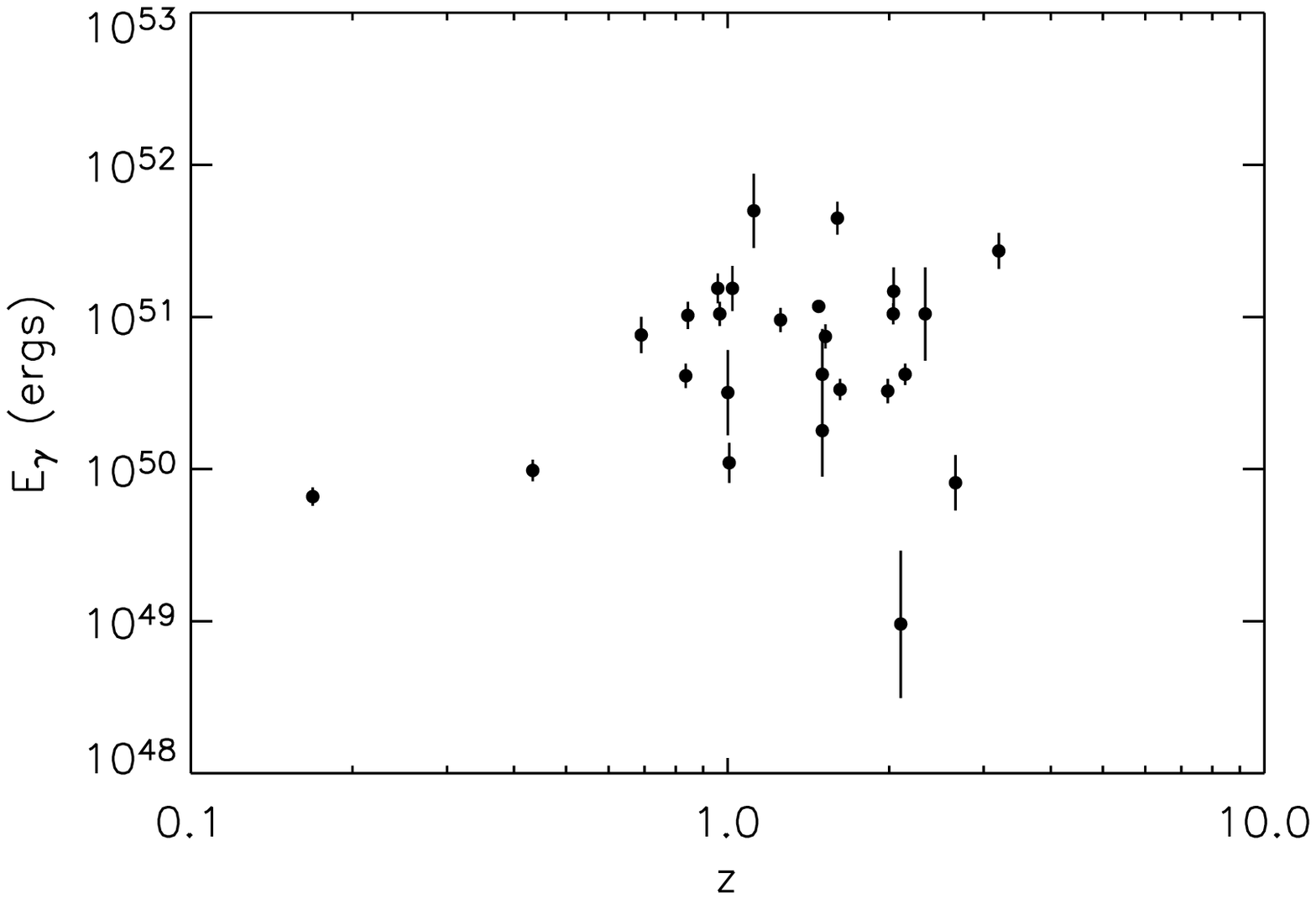}
\end{figure}

\begin{figure} \label{fig:beamingangle}
\plotone{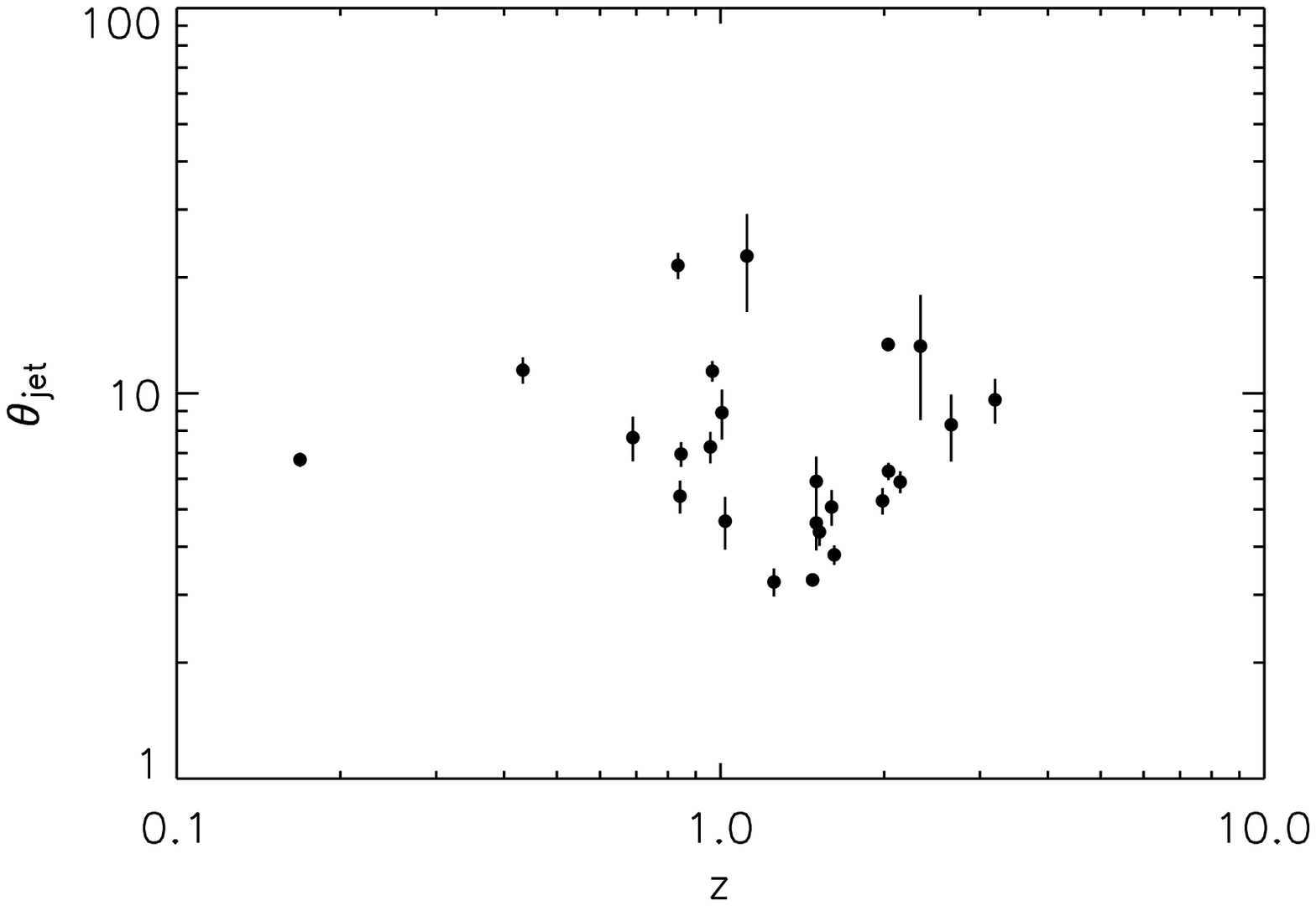}
\end{figure}

Plotted in Figure \ref{fig:fluencevsredshift} are the $E_{\gamma}$ estimates
from \citet{Friedman05} for a little over two dozen GRBs.  Furthermore, plotted in Figure \ref{fig:beamingangle} is the canonical jet opening angle for the same two dozen GRBs.  By apply a standard Kendall rank order $\tau$ statistic we can measure the degree of correlation in these two samples in a nonparametric way (i.e. without assuming an underlying correlation type).  We find a correlation strength of $\tau = 0.093$ between $E_{\gamma}$ and redshift and $\tau = 0.163606$ between $\theta_{j}$ and redshift, where a $tau$ of 1 signifies a significant correlation.  Therefore, there is no deduced redshift dependency that would suggest any evolution of the jet opening angle or $E_{\gamma}$ with redshift in the pre-Swift data set.  This lack of redshift dependency stands to be tested in the Swift era as more GRBs with measured jet break times are observed over a broader redshift range, but if it is confirmed then it would imply that the evolution of some jet property other than the collimation factor must be responsible for the brightening of GRBs with redshift.  Speculations on the nature of this evolution are
dependant on the jet model used to explain the emission. The
simplest model assumes a uniform energy distribution per solid angle $\epsilon(\theta)$ across the jet with a sharp drop beyond $\theta_{j}$.  In this scenario, the observed distribution in GRB
energies is directly due to the diversity of jet opening angles, as
is the observed values of the jet break time $t_{j}$.  The lack of
any concrete evidence for an evolution of $\theta_{j}$ with redshift
combined with the observation that the collimation corrected
$E_{\gamma}$ and $L_{\gamma}$ are very narrowly clustered, create
difficulty for the uniform jet model to explain any kind of
evolution in luminosity. One of the observed conditions above would
have to be broken in order to accommodate any such evolution with
this model.

In a structured jet model, the GRB jets are identical having a
quasi-universal shape with a fixed opening angle and a nonuniform
energy distribution per solid angle.  The diversity in the observed
jet break time and isotropic energies would then be a result of
varying viewing angles away from the jet axis $\theta_{v}$.
Furthermore, an observer viewing the GRB at a small $\theta_{v}$
would see an extremely powerful burst, with the observed luminosity
declining as some function of increasing $\theta_{v}$.  A jet
structure with a functional form of $\epsilon(\theta)^{-2}$ is
required to reproduce the observed $t_{j} \propto E_{iso}$, i.e the
\citet{Frail01} and \citet{bloom03} results.  If the requirement of
a narrow $E_{\gamma}$ and $L_{\gamma}$ distribution is broken, then
any power law structure $\theta^{k}$ could still produce the
observed steepening in the afterglow light curve. In this case, the
luminosity evolution would manifest itself not as an narrowing of
$\theta_{j}$ but rather as an overall increase in the normalization
of $\epsilon(\theta_{j})$. Another possibility would be an evolution
of the morphology of $\epsilon(theta)$ as a function of redshift. If
the power law index $\epsilon(\theta_{j}) \sim \theta_{j}$ has
evolved with time, or if $\epsilon(\theta_{j})$ has evolved from a
non-power-law shape (e.g. a Gaussian profile), such that
$\epsilon(\theta_{j})$ varies more slowly with viewing angle, then
there would be a markedly different luminosity distribution at high
redshift.  A third, rather implausible, explanation is a
preferentially small viewing angle $\theta_{v}$ at higher redshift,
although there is no physical reason to think that this is at all
possible. Therefore, it would seem that evidence of luminosity
evolution in the presence of the observation that $E_{\gamma}$ and
$L_{\gamma}$ are narrowly distributed and the lack of any evidence
of an evolution of $\theta_{j}$ with redshift, favors a
quasi-universal jet model over a uniform jet model. This is
primarily due to the inability of the uniform jet model to explain
any kind of luminosity evolution with redshift without a parallel
evolution in the jet opening angle, something that is not currently
observed.

\subsection{Luminosity Functions and Jet Model Discrimination} \label{sec:jetmodels}

Because the energy distribution $\eta(\theta_{j})$ of the structured
jet model is well defined, it can make specific predictions
regarding the GRB luminosity function $\phi(L)$.  In the case of
power-law structured jets $\epsilon(\theta_{j}) \propto
\theta_{j}^{-k}$, resulting in a predicted luminosity function with
a slope of $\gamma = 1 + 2/k$ \citep{zhang02}. The "canonical" $k=2$
model would yield a luminosity function $\propto L^{-2}$
\citep{rossi02}, whereas the quasi-universal gaussian structured jet
model predicts $\propto L^{-1}$ \citep{lloyd04}. Although the
uniform jet model also exhibits a well defined
$\epsilon(\theta_{j})$, it cannot make any firm predictions about
the shape of the GRB luminosity function due to the random variation
$\theta_{j}$.

The luminosity function deduced from our analysis is well fit by a
single power low with an index of $\propto L'^{-1.5}$ for
luminosities below about 10$^{52}$, with a sharp decline for higher
luminosities. This value is intermediate between the expected value
for the $k=2$ power law and gaussian structure models.  An index of
$L'^{-1.5}$ actually predicts a power law model with $k=4$ which is
much steeper than the $k=2$ shape needed to keep $E_{iso}\theta^{2}
\sim$ constant.  It is also possible that a simple gaussian or power
law profile for $\eta(\theta)$ is simply an oversimplification. It
has been pointed out that by \citet{lamb03} that $\eta(\theta)$
would have to fall off steeper than $\theta^{-2}$ at large angles if
the quasi-uniform jet model is to explain the relative numbers of
x-ray flashes to GRBs.  This may explain why so many studies have
found a sharp decline above some break energy possibly indicating a
different value for $k$ at low and high luminosities, i.e large and
small opening angles respectively.  In any case, it would not be
unfeasible to think that the jet morphology is more complicated than
a simple power or gaussian profile, as is suggested by our results.

\section{Conclusions} \label{conclusion:ch3}

In this work we perform demographic studies on a large sample of
luminosity and redshift data found through the use of the
lag-luminosity correlation.  By applying maximum likelihood
techniques, we are able to obtain an estimate of the luminosity
evolution, luminosity function, and density distributions in a way
that accounts for detector selection effects.  We find that there
exists a strong (11.63 $\sigma$) correlation between luminosity and
redshift that can be parameterized as $L(z) = (1+z)^{2.58}$.  The
resulting cumulative $\Phi(L')$ and differential $d\Phi/dL'$
luminosity functions are well fit by double power laws separated by
a break energy of about $10^{52}$ ergs s$^{-1}$, with $d\Phi/dL'$
exhibiting a power law shape of $L^{-1.5}$ below this luminosity.
This value does not immediately discriminate against any proposed
structured jet models, but it may indicate that a more complicated
jet profile is need to explain the observed luminosity function of
GRBs.  The GRB comoving rate density is found to increase as
$\rho_{\gamma}(z) \propto (1+z)^{2.5}$ to a redshift of $z \approx
1$ followed by flattening and eventual decline above $z>6$.
Although, the conversion between $\rho_{\gamma}(z)$ and an estimate
of the SFR cannot be quantitatively made due to the uncertainty
regarding the GRB progenitors and their initial stellar mass
functions, it can be said that $\rho_{\gamma}(z)$ does qualitatively
follow recent photometric estimates of the SFR, as would be expected
from massive short lived progenitors.  We stress that these
conclusions are based on the validity of the lag-luminosity
correlation, which still stands to be confirmed as new redshift data becomes available.  A full
confirmation, and most probably further calibration, of this
distance indicator will have to wait for addition detections with
the Swift spacecraft, which should have the collecting area
necessary to obtain high signal-to-noise energy dependant light
curves from which to measure statistically significant lags. Even
with the slew of directly measured luminosity and redshift data
expected to come from the Swift mission, empirical distance
indicators may still play an important roll in expanding the
available GRB data set through the use of archival BATSE and
BepooSAX data for statistical studies such as the analysis performed
in this work.





\bigskip

\section*{Figure Captions}

\bigskip

{\bf Fig. 1.} - The luminosity and redshift data used in our analysis as
deduced from the lag-luminosity correlation. The dashed line
represents an imposed 0.5 photons cm$^{-2}$ s$^{-1}\ldots$ cut to
the original 1438 bursts analyzed by BNB04, producing the sharp
cutoff in the data.  This leaves a total of 985 bursts with a median
redshift of 1.64. 

\bigskip

{\bf Fig. 2.} - A plot of the lag-luminosity plane for the events under consideration.  The error in the lag and flux measurements are used to determine the uncertainty in the luminosity values which in effect is used as a weight in the maximum likelihood technique that estimates the correlation strength between $L$ and $z$.   

\bigskip

{\bf Fig. 3.} - A representation of the associated sets used in the Lynden-Bell
technique. For each data point with $(L_{i},z_{i})$, the solid line
represents the minimum luminosity or maximum redshift that the burst
could have had and still have been observed. Employing the
Lynden-Bell technique with associated sets defined as $N_{j}$
($L_{j} < L < \infty$, $0 < z < z_{max}(L_{i})$) produces the
cumulative distribution for the vertical axis, whereas the $M_{j}$
associated set produces the distribution for the data represented on
the horizontal axis.  

\bigskip

{\bf Fig. 4.} - Generated luminosity and redshift data used to test the
ability of the Efron $\&$ Petrosian method to estimate the
correlation strength and functional dependence for data of a given
flux cut.

\bigskip

{\bf Fig. 5.} - A histogram of the difference between the known correlation
index and the reconstructed index $(p-q)$ for a large set of generated $(L,z)$
data with arbitrarily imposed flux thresholds. The error in the
method is tightly centered around $p-q=0$.

\bigskip

{\bf Fig. 6.} - The correlation statistic $\tau$ is plotted as a function
of power law index $g_{\alpha}(z)=(1+z)^{\alpha}$ parameterizing the
luminosity evolution. The solid line represents the $\alpha$ index
that minimizes the correlation between $L'$ and $z$. The dotted
lines show the 1 $\sigma$ \emph{statistical} error in the $\alpha$
parameter. 

\bigskip

{\bf Fig. 7.} - The correlation parameter $\alpha$ is plotted as a function of the flux
cut applied to the BNB04 sample.  The optimal $\alpha$ is highly
dependent on the choice of the cut value, showing the importance of
a good understanding of the detector flux threshold.

\bigskip

{\bf Fig. 8.} - The luminosity and redshift data used in our analysis as
deduced from the lag-luminosity correlation. The dashed line
represents an imposed 0.5 photons cm$^{-2}$ s$^{-1}\ldots$ cut to
the original 1438 bursts analyzed by BNB04, producing the sharp
cutoff in the data.  This leaves a total of 985 bursts with a median
redshift of 1.64.

\bigskip

{\bf Fig. 9.} - The present epoch GRB luminosity function $\psi(L') =
d\Phi/dL'$, representing the number of events between the luminosity
$L'$ and $L'+dL'$.  The function falls as roughly $\propto
L'^{-1.5}$ for luminosities below the break energy of $10^{52}$ ergs
s$^{-1}$. 

\bigskip

{\bf Fig. 10.} - The cumulative density distribution
$\sigma(z)=\int_{0}^{z}\rho(z)(dV/dz)dz$, representing the total
number of events up to a given redshift.  The flattening at high
redshift indicates a drop off of events around $z \sim 5-10$.

\bigskip

{\bf Fig. 11.} - The
cumulative density function $\sigma(z)$ plotted as a function of
comoving volume.  The dashed line represents the increase in the
number of sources if the GRB density were constant throughout the
history of the universe.  The GRB density has increased as $\rho
\sim V^{0.25}$ before peaking between $z \sim 1-2$ and then
decreasing.

\bigskip

{\bf Fig. 12.} - The comoving rate density $\rho(z)$ as a function of
redshift.  The rate density of sources can be seen to follow the
evolution deduced from Figure \ref{fig:cumdensityvolume}, increasing to a
redshift of 1-2 then flattening before decreasing at higher
redshifts.  The circles represent high redshift cc rates
from \citet{dahlen04} whereas the square point is the local rate
found \citet{capp99}.  The increase in the cc event rate with
redshift qualitatively matches the overall increase in the GRB
comoving rate density.
 
\bigskip

{\bf Fig.  13.} - Plot of the collimation corrected total emitted energy of
23 GRBs with known redshift and beaming angles.  No significant
correlation can be seen with redshift.

\bigskip

{\bf Fig. 14.} - The beaming angles $\theta_{j}$ of 23 GRBs with known
redshift.  The lack of a significant correlation with redshift is
quite evident.

\bigskip


\section*{Figures}

\begin{thebibliography}{}

\bibitem[Amati et al.(2002)]{amati02} Amati, L., et al. 2002, A$\&$A 390, 81
\bibitem[Band, Norris, Bonnell(2004)]{band04} Band, D., Norris, J., Bonnell, J. 2004 ApJ, 613, 484
\bibitem[Band et al.(1993)]{band93b} Band, D., et al. 1993, AAS, 182, 7409B
\bibitem[Band $\&$ Preece(2005)]{bandpreece} Band, D., $\&$ Preece, R. 2005 ApJ, 627, 319
\bibitem[Bloom(2003)]{bloom03b} Bloom, J. S., 2003, PASP, 115, 271
\bibitem[Bloom et al.(2003)]{bloom03} Bloom, J. S., Frail, D., $\&$ Kulkarni, S. 2003b, ApJ, 594, 678
\bibitem[Bouwens et al.(2003)]{Bouwens03} Bouwens et. al. 2003, ApJ, 595, 589B
\bibitem[Boyle et al.(1993)]{Boyle93} Boyle et. al. 1993, MNRAS, 260, 49B
\bibitem[Cappallaro et al.(1999)]{capp99} Cappellaro, E., Evans, R., $\&$ Turatto, M. 1999, A$\&$A, 351, 459
\bibitem[Dahlen et al.(2004)]{dahlen04} Dahlen, T., 2004, ApJ, 613, 189D
\bibitem[Djorgovski et al.(1998)]{Djorg98} Djorgovski, S., Kulkarni, S., Bloom, J., Goodrich, R., Frail, D., Piro, L., Palazzi, E., 1998, ApJ, 508L, 17D
\bibitem[Efron $\&$ Petrosian(1992)]{efron92} Efron, B., Petrosian, V. 1992, ApJ, 399, 345
\bibitem[Fall, Charlot, $\&$ Pei(1996)]{fall96} Fall, S., Charlot, S., Pei, Y. 1996, ApJ, 464L, 43F
\bibitem[Fishman et al.(1994)]{fish94} Fishman, G. J., et. al. 1994, ApJS, 92, 229
\bibitem[Fontana et al.(2003)]{Fontana03} Fontana et. al. 2003, ApJ, 587, 544F
\bibitem[Frail et al.(2001)]{Frail01} Frail, D. A., et. al. 2001, ApJ, 562, L55
\bibitem[Friedman $\&$ Bloom(2005)]{Friedman05} Friedman, A., Bloom, J. 2005, ApJ 627, 1
\bibitem[Ghirlanda et al.(2004a)]{ghirlanda04a} Ghirlanda, G., Ghisellini, G. and Lazzati, D. 2004, ApJ, 616, 331G
\bibitem[Guetta et al.(2003)]{guetta03} Guetta, D., Piran, T., Waxman, E., 2005 ApJ, 619, 412G
\bibitem[Hjorth et al.(2003)]{hjorth} Hjorth, J. et al. 2003, Nature 423, 847
\bibitem[Ioka $\&$ Nakamura(2001)]{Ioka01} Ioka, K., Nakamura, T., 2001, ApJ, 554L, 163I
\bibitem[Iwata et al.(2003)]{Iwata03} Iwata, I. 2003, PASJ, 55, 415I
\bibitem[Kobayashi, Ryde, $\&$ MacFadyen(2002)]{krm02} Kobayashi, S., Ryde, F., MacFadyen, A. 2002, ApJ, 577, 302
\bibitem[Kocevski $\&$ Liang(2003)]{kocevski03a} Kocevski, D., E. P. Liang 2003, ApJ 594, 385K
\bibitem[Kulkarni et al.(1998)]{Kulkarni98} Kulkarni, S. R. et al. 1998, Nature 393, 35 
\bibitem[Lamb $\&$ Graziani(2003)]{lamb03} Lamb, D.,Graziani, C. 2003, AAS, 202, 4501L
\bibitem[Lloyd-Ronning, Fryer, $\&$ Ramirez-Ruiz(2002)]{lloyd02} Lloyd-Ronning, N., Fryer, C., Ramirez-Ruiz, E., 2002, ApJ, 574, 554L
\bibitem[Lloyd-Ronning, Dai, $\&$ Zhang(2004)]{lloyd04} Lloyd-Ronning, N., Dai, X., Zhang, B., 2004, ApJ, 601, 371
\bibitem[Lynden-Bell(1971)]{Lynden-Bell71} Lynden-Bell, D., 1973 Mon. Not. R. astr. soc., 155, 95
\bibitem[MacFadyen $\&$ Woosley (1999)]{macfadyen99}  MacFadyen, A. I. $\&$ Woosley, S. E. 1999, ApJ 524, 262
\bibitem[Madau et al.(1996)]{madau96} Madau et. al. 1996, MNRAS, 283, 1388M
\bibitem[Maloney $\&$ Petrosian(1999)]{Maloney99} Maloney, A., Petrosian, V. 1999, ApJ, 518, 32
\bibitem[Murakami et al.(2003)]{Murakami} Murakami et al. 2003, PASJ, 55, L65
\bibitem[Nakar $\&$ Piran (2005)]{nakar} Nakar, E. $\&$ Piran, T.  2005, MNRAS, 360L, 73N
\bibitem[Norris, Marani, $\&$ Bonnell(2000)]{norris00} Norris, J. P., Marani, G. F. $\&$ Bonnell, J. T. 2000, Astrophys. J 534, 248
\bibitem[Norris(2002)]{norris02} Norris, J., 2002, ApJ, 579, 386N
\bibitem[Petrosian(1992)]{Petrosian92} Petrosian, V. 1992 in Statistical Challenges in Modern Astronomy, ed. E. D. Feigelson $\&$ G. J. Babu (New York: Springer), 173
\bibitem[Riess et al.(2004)]{riess04} Riess et. al. 2004, ApJ, 607, 665R
\bibitem[Rossi, Lazzati, $\&$ Rees(2002)]{rossi02} Rossi, E., Lazzati, D., $\&$ Rees, M. 2002 MNRAS, 332, 945R
\bibitem[Salmonson (2000)]{salmonson00} Salmonson, J. D. 2000, ApJ, 544, L115
\bibitem[Salmonson (2001)]{salmonson01} Salmonson, J. D. 2001, AAS, 198, 3809S
\bibitem[Sazonov, Lutovinov, $\&$ Sunyaev(2004)]{Sazonov04} Sazonov, S., Lutovinov, A., Sunyaev, R. 2004, Nature, 430, 646S
\bibitem[Schaefer, Deng, $\&$ Band]{Schaefer01} Schaefer, B., Deng, M., Band, D. 2001, ApJ, 563L, 123S
\bibitem[Schmidt(2001)]{Schmidt01} Schmidt, M. 2001, ApJ, 552,36S
\bibitem[Soderberg(2004)]{Soderberg04} Soderberg, A. 2004, Nature, 430, 648S
\bibitem[Stanek et al.(2003)]{stanek03} Stanek, K. Z. et al. 2003, ApJ 591, L17
\bibitem[Steidel et al.(1999)]{Steidel99} Steidel et. al. 1999, ApJ, 519, 1S
\bibitem[Yonetoku et al.(2004)]{Yonetoku04} Yonetoku, D. 2004, ApJ, 609, 935Y
\bibitem[Woosley(2000)]{woosley00} Woosley, S. E. 2000, AIPC, 526, 555W
\bibitem[Zhang $\&$ M\'{e}sz\'{a}ros(2002)]{zhang02} Zhang, B. $\&$ Meszaros, P. 2002, ApJ 571, 876

\end{thebibliography}
\end{document}